\documentclass[12pt]{article}
\usepackage[footnotesize]{caption}
\usepackage{geometry}
\usepackage{hyperref}
\usepackage{amsmath,amsfonts,amssymb,latexsym}
\usepackage{graphicx,subfig}
\usepackage{color}
\usepackage{cite}

\numberwithin{equation}{section} 
\numberwithin{figure}{section}
\numberwithin{table}{section} 
\newcommand{\dev}[3][]{\frac{\mathrm{d}^{#1} #2}{\mathrm{d} #3^{#1}}}
\newcommand{\pdev}[3][]{\frac{\partial^{#1} #2}{\partial #3^{#1}}}

\begin{document}

\begin{titlepage}
\begin{center}

{\Large {\bf Universality for quintessence.}}
 \\
 ~\\
\vskip 2cm

\vskip 1cm

{\bf  \large F. Cicciarella${}^{a,}$\footnote{f.cicciarella1@gmail.com}, M. Pieroni${}^{b,c,}$\footnote{mauro.pieroni@apc.in2p3.fr}. }\\
~\\
~\\
{${}^{a}$\em Dipartimento di Fisica, Universit\`a di Pisa, Largo Bruno Pontecorvo 3, 56127 Pisa}\\
{${}^{b}$\em Laboratoire AstroParticule et Cosmologie,
Universit\'e Paris Diderot } \\
{${}^{c}$\em Paris Centre for Cosmological Physics, F75205 Paris Cedex 13}\\

\end{center}

\vskip 1cm
\centerline{ {\bf Abstract}}{ Several recent works suggested the possibility of describing inflation by means of a renormalization group equation. In this paper we discuss the application of these methods to models of quintessence. In this framework a period of exponential expansion corresponds to the slow evolution of the scalar field in the neighborhood of a fixed point. A minimal set of universality classes for models of quintessence is defined and the transition from a matter dominated to quintessence dominated universe is studied. Models in which quintessence is non-minimally coupled with gravity are also discussed. We show that the formalism proves to be extremely convenient to describe quintessence and moreover we find that in most of the models discussed in this work quintessence naturally takes over ordinary matter.}

\vskip .5cm
\indent

\vfill

\end{titlepage}

\newpage

\tableofcontents

\section{Introduction.\label{sec:introduction}}
According to present observations such as the mapping of cosmic microwave background (CMB) anisotropies performed by the Planck satellite~\cite{Ade:2015xua}, the standard cosmological model (typically referred to as $\Lambda$CDM) supplemented with inflation appears to be in remarkably good agreement with almost all data. In this model the presently observed accelerated expansion of the Universe due to the presence of Dark Energy (DE) is modeled by the introduction of a Cosmological Constant (CC), typically denoted with $\Lambda$. The CC is thus playing the role of an energy species with negative pressure $p_{\Lambda}=-\rho_{\Lambda}$. The measured value of $\rho_{\Lambda_{\mathrm{vac}}}$ in reduced Planck units, \emph{i.e.} as a natural dimensionless value, is $\simeq 10^{-122}$. The absence of theoretical reasons to justify such an absurdly small value for $\rho_{\Lambda}$ results in a so-called fine-tuning problem. For this and other reasons (such as the coincidence problem), the case of a CC is not completely satisfactory from a theoretical point of view. \\

Over the years several alternatives to the simplest scenario were proposed. One of the simplest realizations of a dynamical DE is obtained by considering a minimally coupled homogeneous scalar field $\phi$ which evolves in a given potential\footnote{An homogeneous Universe with an effective CC is the late time attractor of the generic inhomogeneous case with an effective CC~\cite{Starobinsky:1982mr}.}. Such a scenario, which was originally introduced in~\cite{Ratra:1987rm,Wetterich:1987fm,Peebles:1987ek,Caldwell:1997ii,Zlatev:1998tr}, is typically referred to as `quintessence'. Among the other possibilities it is worth mentioning theories of modified gravity such as $f(R)$ theories~\cite{Capozziello:2002rd,Capozziello:2003tk} and scalar-tensor theories\footnote{These theories are typically based on generalizations of the original Brans-Dicke theory~\cite{Brans:1961sx}. The formulation of the most general scalar field theories with an action depending on derivatives up to second order with second order equations of motion (recently rediscovered in~\cite{Deffayet:2011gz,Kobayashi:2011nu}) is due to Horndeski~\cite{Horndeski:1974wa}.} where a scalar field has a generalized coupling with gravity~\cite{Cooper:1982du,Finelli:2007wb,Uzan:1999ch,Chiba:1999wt,Amendola:1999qq,Perrotta:1999am,Bertolami:1999dp,Boisseau:2000pr,EspositoFarese:2000ij}. For a comprehensive review of models for DE and modified gravity see for example~\cite{Amendola:2012ys}. \\

It is known in the literature that in the case of minimally coupled DE models attractor solutions may exist~\cite{Ratra:1987rm,Copeland:1997et,Liddle:1998xm,Steinhardt:1999nw}. If an attractor exists\footnote{The existence of attractors depends only on few characteristics of the potential.}, the evolution of several models corresponding to a wide set of different initial conditions converges towards a unique asymptotic behavior. Interestingly this universality is not only achieved during the late time evolution if the Universe is filled by a homogeneous scalar field, but it is also present at early times during inflation\footnote{Nevertheless, it is crucial to stress that in the case of inflation we depart from the attractor configuration while in the case of quintessence we approach this configuration.}~\cite{Binetruy:2014zya,Pieroni:2016gdg,Pieroni:2015cma,Binetruy:2016hna}. Indeed, the presence of universality is due to the fact that in these regimes the dynamics depends only on few characteristic features of the potentials for the scalar fields that drive the acceleration. \\

In the case of inflation an interesting explanation for this universality arises if we consider the possibility of describing the evolution of the inflating Universe in terms of a Renormalization Group (RG) equation. Such a description is based on the application of the Hamilton-Jacobi formalism of Salopek and Bond~\cite{Salopek:1990jq} to describe the cosmological evolution of a scalar field in its potential\footnote{The Hamilton-Jacobi formalism to describe the evolution of a FRLW spatially flat universe filled by a minimally coupled scalar field was independently introduced in~\cite{Muslimov:1990be}.}. In this context inflation is interpreted as the slow evolution of a system leaving a critical fixed point which corresponds to the asymptotic de Sitter (dS) configuration. As a consequence it is possible to use the Wilsonian picture of fixed points (exact dS solutions), scaling regions (periods of accelerated expansion), and critical exponents (scaling exponents of the spectra) to give a universal characterization of inflation~\cite{Binetruy:2014zya,Pieroni:2016gdg}. Indeed the possibility of describing inflation as an RG flow is not fortuitous. In particular, by performing an analytical continuation it is possible to map cosmological solutions that asymptote to dS into domain-wall solutions that asymptote to Anti de Sitter (AdS). An holographic description (using the AdS/CFT correspondence of Maldacena~\cite{Maldacena:1997re}) of inflation can thus be carried out. In this context inflation corresponds to the RG flow of the dual (3-dimensional) quantum field theory (QFT)~\cite{McFadden:2009fg,McFadden:2010na,Kiritsis:2013gia,Garriga:2014fda,Garriga:2015tea,Garriga:2016poh}. \\

By analogy with the case of inflation, it seems reasonable to consider the possibility of describing the accelerated expansion due to DE as an RG flow towards an attractive fixed point in the future. In particular, using the $\beta$-function formalism developed for inflation to describe DE, we can easily identify the asymptotic behavior associated with models that belong to different universality classes. Once the asymptotic behavior is specified, the transition from a matter dominated phase to a phase where the evolution is dominated by quintessence can be studied. In particular, in this work we show that once the constraints set by present time observations are imposed, this transition naturally occurs for a wide class of models. \\

In this paper we proceed as follows: In Sec.~\ref{sec:formalism} we present the $\beta$-function formalism for inflation introduced in~\cite{Binetruy:2014zya} and we discuss its application to describe models for quintessence. In particular we analyze the asymptotic behavior and in Sec.~\ref{sec:univ-quint} we introduce a set of universality classes for models of quintessence. In Sec.~\ref{sec:formalism_matter} we generalize the formalism in order to discuss the cosmological evolution of a Universe that is filled by both matter and quintessence. In Sec.~\ref{sec:formalism_non_min} we consider the possibility of having a non-minimal coupling between quintessence and gravity and finally in Sec.~\ref{sec:conclusions} we draw our conclusions.

\section{The $\beta$-function formalism. \label{sec:formalism}}
The action for a homogeneous scalar field $\phi(t)$ with potential $V(\phi)$ is:
\begin{equation}
\mathcal{S}=\int\mathrm{d}^4x\;\sqrt{-g}\left[m_p^2\frac{R}{2}-\frac{1}{2}g^{\mu\nu}\partial_{\mu}\phi\partial_{\nu}\phi-V(\phi)\right] \ , \label{inflation}
\end{equation}
where $m_p^2 \equiv M_p^2/(8 \pi) =  1/(8 \pi G_N)$. In the case of a flat Friedmann-Lema\^itre-Robertson-Walker (FLRW) universe with $(-,+,+,+)$ signature:
\begin{equation}
\mathrm{d}s^2=-\mathrm{d}t^2+a^2(t)\mathrm{d}\mathbf{x}^2 \ ,
\end{equation}
the only two independent Einstein equations can be expressed as:
\begin{align}
H^2\equiv \left(\frac{\dot{a}}{a}\right)^2 &= \frac{\rho_{\phi}}{3 \, m_p^2} \ , \label{fried1} \\
-2\dot{H} &= \frac{p_{\phi} + \rho_{\phi}}{m_p^2}\ , \label{fried2}
\end{align}
where a dot is used to denote a derivative with respect to cosmic time and, according to the definition of the energy-momentum tensor $T_{\mu\nu}$:
\begin{equation}
T_{\mu\nu}\equiv -\frac{2}{\sqrt{-g}}\frac{\delta S}{\delta g^{\mu\nu}}=\partial_{\mu}\phi\partial_{\nu}\phi-g_{\mu\nu}\left(\frac{1}{2}g^{\alpha\beta}\partial_{\alpha}\phi\partial_{\beta}\phi+V(\phi)\right) \ ,
\end{equation}
we have defined:
\begin{equation}
\label{infldens_press} 
\rho_{\phi} \equiv \frac{\dot{\phi}^2}{2}+V(\phi) \ , \qquad \qquad p_{\phi} \equiv \frac{\dot{\phi}^2}{2}-V(\phi)\ , 
\end{equation}
as the energy density and the pressure associated with the scalar field $\phi$. Notice that as usual a phase of nearly exponential expansion (\emph{i.e.} $H$ nearly constant) is realized if $p_{\phi} \simeq -\rho_{\phi}$ (\emph{i.e.} if $\dot{\phi}^2 \ll V(\phi)$).  \\

By varying the action~\eqref{inflation} with respect to $\phi$, we obtain the equation of motion for the scalar field, namely:
\begin{equation}
\ddot{\phi}+3H\dot{\phi}=-\dev{V}{\phi} \ , \label{eom}
\end{equation}
which is equivalent to the continuity equation for the energy density:
\begin{equation}
\dot{\rho_{\phi}}=-3H(p_{\phi}+\rho_{\phi}) \ , \label{continuity}
\end{equation}
that directly follows from the conservation law for the energy-momentum tensor $\nabla^{\mu}T_{\mu\nu}=0$.\\

Following the approach of Salopek and Bond~\cite{Salopek:1990jq}, we proceed by assuming that the evolution of the scalar field as a function of time is - at least - piecewise monotonic. Under this assumption and by inverting the dependence $\phi(t)$ into $t(\phi)$, the field itself provides a clock. Hence, in this framework the Hubble parameter may be regarded as a function of $\phi$ and it is convenient to define:
\begin{equation}
\label{superpot_def}
H=\frac{\dot{a}}{a}\equiv -\frac{1}{2}W(\phi) \ .
\end{equation}
In terms of $W$, the equations for $H$ and its time derivative can be expressed as (using\footnote{We use the notation $W_{,\phi}\equiv \mathrm{d}W/\mathrm{d}\phi$.} $\dot{H}=-W_{,\phi}\dot{\phi}/2$):
\begin{align}
\frac{3}{4}m_p^2W^2(\phi) \quad &=\quad  \rho_{\phi} \quad= \quad\frac{\dot{\phi}^2}{2}+V(\phi), \label{pot_HJ} \\
-2m_p^2\dot{H} \quad= \quad m_p^2W_{\phi}\dot{\phi} \quad &=\quad p_{\phi}+\rho_{\phi}\quad=\quad\dot{\phi}^2 \ . \label{dotH}
\end{align}
Notice that from Eq.\eqref{dotH} we immediately obtain $\dot{\phi}=m_p^2W_{,\phi}$, implying that $\dot{\phi}$ can be expressed as a function of $\phi$ only. Substituting into Eq.~\eqref{pot_HJ} we can then express the inflaton potential as:
\begin{equation}
2\frac{V}{m_p^2}=\frac{3}{2}W^2-m_p^2 W_{,\phi}^2 \ . \label{pot-superpot}
\end{equation}
By analogy with a similar parametrization of the potential in the context of supersymmetric quantum mechanics (see for example~\cite{Binetruy:2006ad}) we are thus led to call the function $W(\phi)$ \emph{superpotential}. For completeness we express the pressure and the energy density in terms of the superpotential as:
\begin{align}
\rho_{\phi}\quad & = \quad \frac{3}{4} m_p^2 W^2 \ , \label{friedmann1} \\
\rho_{\phi}+p_{\phi}\quad &= \quad m_p^4 W_{,\phi}^2 \ .\label{friedmann2}
\end{align}
At this point it is interesting to notice that defining (from now on, we set $m_p^2=1$):
\begin{equation}
 \label{betafunction}
	\beta(\phi)\equiv \dev{\phi}{\ln a}= \frac{\dot{\phi}}{H} =- 2 \, \frac{W_{,\phi}}{W}
\end{equation}
the cosmological evolution of the scalar field $\phi$ in its potential can be expressed in terms of a renormalization group equation (RGE) where the $\phi$ plays the role of the coupling constant $g$ and the scale factor $a$ plays the role of the renormalization scale $\mu$. In this framework fixed points then correspond to zeros of $W_{,\phi}$, which are also extrema of the potential $V(\phi)$. \\

As discussed in~\cite{Binetruy:2014zya,Pieroni:2016gdg}, by fixing a parameterization for $\beta$ we specify the evolution of the system (or equivalently the RG flow) in a neighborhood of a fixed point (i.e. in a region where $\beta(\phi)\ll 1$). As a single asymptotic behavior can be reached (by neglecting higher order terms) by several models, it is natural to regroup models into a set of universality classes. For this reason, by specifying a parameterization for the $\beta$-function we are not simply specifying a single model but rather a set of theories sharing a scale invariant limit.

\subsection{Universality for Quintessence.}
\label{sec:univ-quint}
In order to describe quintessence within the $\beta$-function formalism, we can start by expressing the superpotential and the potential associated with a given parameterization of the $\beta$-function:
\begin{align}
W(\phi)\quad &=\quad W_0\exp\left\{-\frac{1}{2}\int_{\phi_0}^{\phi}\mathrm{d}\phi'\;\beta(\phi')\right\} \ , \label{superpot_expr}\\
V(\phi)\quad &=\quad \frac{3}{4}W_0^2\exp\left\{-\int_{\phi_0}^{\phi}\mathrm{d}\phi'\;\beta(\phi')\right\}\left(1-\frac{\beta^2(\phi)}{6}\right) \ ,\label{potential_expr} 
\end{align}
where $W_0$ and $\phi_0$ are respectively used to denote the value of the superpotential and the value of $\phi$ at present time\footnote{Notice that these expression (and most of the expressions shown in this section) are only valid in the limit that the contribution of other fluids (matter, radiation, ...) to the evolution is negligible. The generalization of these expressions to include the presence of other energy species is discussed in Sec.~\ref{sec:formalism_matter}.}. In order to reproduce a nearly-exponential expansion we must satisfy the constraint:
\begin{equation}
\frac{p_{\phi}+\rho_{\phi}}{\rho_{\phi}}=\frac{4}{3}\frac{W_{,\phi}^2}{W^2}=\frac{\beta^2(\phi)}{3}\rightarrow 0 \ , \label{eqofstate}
\end{equation}
as $\phi$ approaches the fixed point. Let us recall that the equation of state parameter $w_{\phi}$ associated with an energy species is defined as $p_{\phi} \equiv w_{\phi} \rho_{\phi}$. As a consequence we directly get:
\begin{equation}
	\label{parameter_beta_definition}
	1 + w_{\phi} = \frac{p_{\phi}+\rho_{\phi}}{\rho_{\phi}} = \frac{\beta^2(\phi)}{3} \ .
\end{equation}
In the literature, the evolution of $1+w_{\phi}$ is commonly showed in terms of the redshift $z$ defined as $1+z=a_0/a$. Taking the logarithm of this definition and differentiating we obtain $\mathrm{d}z/(1+z)=-\mathrm{d} \ln a$ so that we get:
\begin{equation}
	\label{redshift_beta}
\beta(\phi) = \dev{\phi}{\ln a}=-(1+z)\dev{\phi}{z} \qquad \rightarrow \qquad \ln(1+z) = -\int_{\phi_0}^{\phi}\frac{\mathrm{d}\phi'}{\beta(\phi')} .
\end{equation}
Solving this equation for $\phi$ we can thus express $w_{\phi}$ as a function of $z$.\\

In order to justify the small value of $\rho_{\Lambda_{\mathrm{vac}}}$ without tuning the parameters of the model, we have to realize $\rho_{\phi}\to 0$ dynamically. Combining Eq.~\eqref{superpot_expr} and Eq.~\eqref{friedmann1} we can see that, as $\phi$ approaches the fixed point, this request translates into:
\begin{equation}
\rho_{\phi}\rightarrow 0 \ , \qquad \iff \qquad \int_{\phi_0}^{\phi}\mathrm{d}\phi'\; \beta(\phi') \rightarrow +\infty \ . \label{0energy}
\end{equation}
As Eq.~\eqref{eqofstate} imposes that approaching the fixed point $\beta \rightarrow 0$ and Eq.~\eqref{0energy} requires its integral to be infinite, we conclude that (assuming $\beta(\phi)$ to be continuous) the fixed point has to be reached for $\left| \phi \right| \rightarrow \infty$. For simplicity, we proceed by assuming that the fixed point is approached for $\phi \rightarrow +\infty $. Notice that combining Eq.~\eqref{friedmann2} with Eq.~\eqref{betafunction} we get:
\begin{equation}
	\label{beta_sign}
 	\textrm{d}t = - 2 \frac{\textrm{d}\phi}{\beta(\phi) W(\phi)}  \ .
 \end{equation}  
As $W = -2H < 0$ and (as the fixed point is approached for $\phi \rightarrow +\infty $) $\textrm{d}\phi > 0$, Eq.~\eqref{beta_sign} implies that $\beta(\phi)>0$ and $\beta_{,\phi}(\phi)<0$.\\

At this point we can also impose:
\begin{equation}
 \label{infinitetime}
\int_{\phi_0}^{\phi} \frac{\mathrm{d}\phi'}{\beta(\phi')} \rightarrow +\infty  \ ,
\end{equation}
\emph{i.e.} we require that the fixed point of the $\beta$-function is reached in the infinite future\footnote{To be more accurate the fixed point is reached in the infinite future if:
\begin{equation}
	\label{t_infinite}
	t - t_0 = \int_{\phi_0}^{\phi} 	\frac{\textrm{d}\hat{\phi}}{W_{,\hat{\phi}} (\hat{\phi})} = - 2 \int_{\phi_0}^{\phi} \frac{\textrm{d}\hat{\phi}}{\beta(\hat{\phi}) \, W(\hat{\phi})} \rightarrow \infty \ .
\end{equation}
As Eq.~\eqref{0energy} implies $W(\phi)\rightarrow 0$, Eq.~\eqref{t_infinite} sets a condition on $\beta(\phi)$ which is less stringent than the condition set by Eq.~\eqref{infinitetime}. Indeed Eq.~\eqref{infinitetime} is not only implying that the fixed point is approached in the infinite future but it is also implying that the scale factor becomes infinitely large approaching the fixed point.}. Any parametrization for $\beta(\phi)$ that is simultaneously satisfying the constraints set by Eq.~\eqref{eqofstate}, Eq.~\eqref{0energy} and Eq.~\eqref{infinitetime} is thus defining a cosmological solution which can be suitable to describe quintessence. \\ 

As we are interested in considering some explicit examples of the asymptotic expression for $\beta(\phi)$, we start by noticing that any function respecting:
\begin{equation}
\lim_{\phi \rightarrow \infty}  \frac{1}{\phi}  \lesssim \lim_{\phi \rightarrow \infty}  \beta(\phi) = 0 \ ,
\end{equation}
is simultaneously satisfying the constraints set by Eq.~\eqref{eqofstate}, Eq.~\eqref{0energy} and Eq.~\eqref{infinitetime}. As a consequence we consider\footnote{In the following we use $f(x) = o(g(x))$ to denote:
\begin{equation}
\lim_{x \rightarrow \infty} f(x)/g(x) \rightarrow 0 \ .	
\end{equation}
This notation is similar to the so called ``little-o notation'' but we are not requiring $\lim_{x \rightarrow \infty} g(x) \neq 0$.}:
\begin{itemize}
\item \textbf{Inverse monomial class}: $\beta(\phi) = \beta_1/\phi \, + o(1/\phi)$ with $\beta_1 > 0$. Using Eq.~\eqref{potential_expr} we can compute the asymptotic expression for $V(\phi)$:
\begin{equation}
\label{inverse_monom_pot}
V(\phi)=\frac{3}{4}W_0^2\left(\frac{\phi_0}{\phi}\right)^{\beta_1}\left(1-\frac{\beta_1^2}{6\phi^2}\right)\simeq \frac{3}{4}W_0^2\left(\frac{\phi_0}{\phi}\right)^{\beta_1}+\mathcal{O}\left(\frac{1}{\phi^{\beta_1+2}}\right) \ .
\end{equation}
By defining the mass scale $M \equiv (3W_0^2\phi_0^{\beta_1}/4)^{1/(4+\beta_1)}$, the asymptotic expression of the potential can thus be cast in the form:
\begin{equation}
V(\phi)\simeq M^{4+\beta_1}\phi^{-\beta_1} \ ,
\end{equation}
which corresponds  to the Ratra-Peebles ``tracking" potential discussed in~\cite{Peebles:1987ek,Zlatev:1998tr,Steinhardt:1999nw}.
\item \textbf{Inverse fractional class}: $\beta(\phi) = \beta_{\alpha}/\phi^{\alpha} \, + o(1/\phi^{\alpha})$ with $0<\alpha<1$ and $\beta_\alpha > 0$. In this case, the potential reads:
\begin{equation}
V(\phi)=\frac{3}{4} W_0^2 \exp\left\{- \beta_\alpha \frac{\phi^{1-\alpha}- \phi_0^{1-\alpha}}{1-\alpha}\right\} \left(1-\frac{\beta_{\alpha}^2}{6 \phi^{2\alpha}}\right) \ .
\end{equation}
\item \textbf{Inverse logarithmic class}: $\beta(\phi) = \beta_0/(\phi\ln\phi) \, + o(1/\phi\ln\phi) $ with $\beta_0 > 0 $. In this case, the potential is given by:
\begin{equation}
V(\phi)=\frac{3}{4} W_0^2 \left( \frac{\ln \phi_0}{ \ln \phi} \right)^{\frac{\beta_0}{2}} \left( 1- \frac{\beta_0^2}{6 \phi^2 \ln^2 \phi }\right) \ .
\end{equation}
\item \textbf{Asymptotic power-law}: $\beta(\phi) = \gamma+f(\phi)$ with $\gamma > 0$ and $f(\phi\to\infty)\to 0$. In order to be consistent with our requests (in particular with Eq.~\eqref{eqofstate}), we also have to impose\footnote{\label{asymptotic_footnote}As discussed in~\cite{Binetruy:2014zya,Pieroni:2016gdg} this particular parametrization for $\beta(\phi)$ does not lead to an asymptotic dS solution but rather to an asymptotic power-law solution and the holographic 3-dimensional QFT is not a CFT.} $\gamma\ll 1$. Notice that a strictly constant $\beta$-function is forbidden, since going back in time there is no exit from the phase of accelerated expansion\footnote{This conclusion changes if (cold dark) matter is taken into account. For more details see Sec.~\ref{sec:formalism_matter} and Appendix~\ref{sec:appendix_other} (in particular the case of a constant $\beta$-function is presented in Appendix~\ref{sec:appendix_power_law}).}. The asymptotic expression for the potential is:
\begin{equation}
V(\phi)\simeq V_0 e^{-\gamma(\phi-\phi_0)} \ .
\end{equation}
This family of exponential potentials has been discussed in~\cite{Capozziello:2005ra,Rubano:2001su,Rubano:2003et,Pavlov:2001dt,Rubano:2002mc,Demianski:2004qt}. It is also interesting to notice that this expression for the $\beta$-function only requires weak constraints on $f(\phi)$. In particular, since the presence of the small constant term makes both the integrals of Eq.~\eqref{0energy} and Eq.~\eqref{infinitetime} divergent, we only need $f(\phi) \rightarrow 0$ approaching the fixed point. For example we can choose $f(\phi)$ to be one of the functions discussed above\footnote{In this case, using the same mechanism as described in~\cite{Binetruy:2014zya}, it is possible to interpolate between the asymptotic power-law class and other classes presented in this section.} or alternatively we can choose a new parametrization for $f(\phi)$ (for example $f(\phi)=e^{-\delta\phi}$ with $\delta>0$).
\end{itemize}

\begin{figure}[h]
	\centering
	\includegraphics[width=0.75\linewidth]{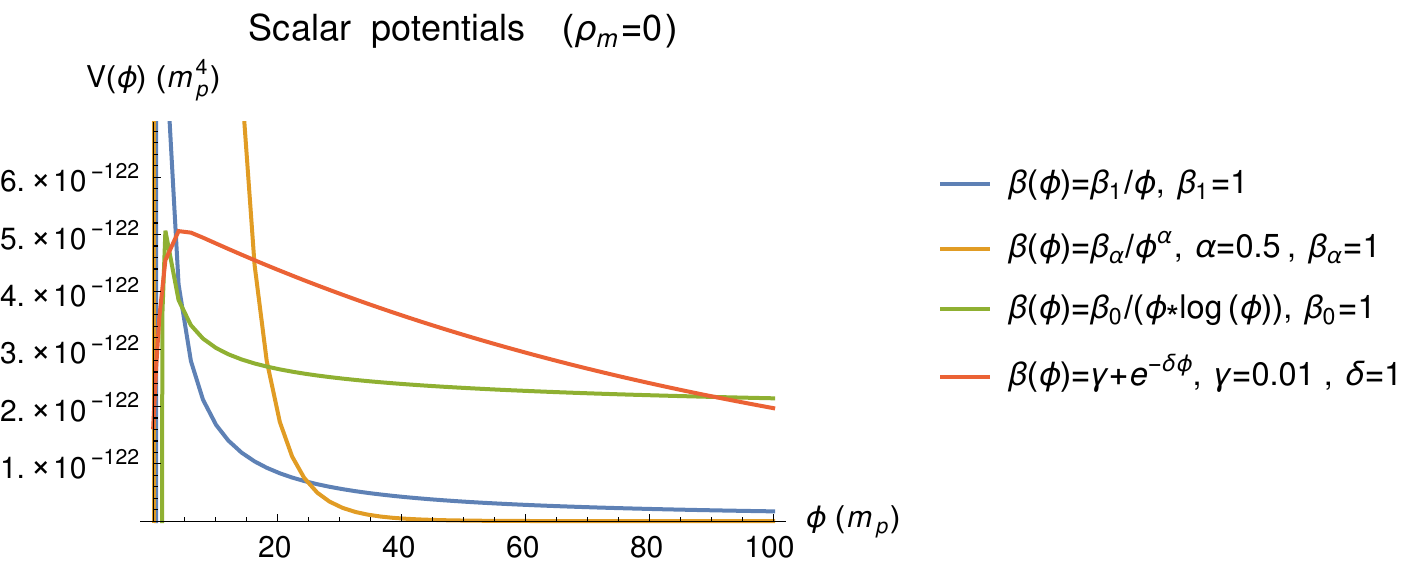}
	\caption{\footnotesize{Profile of the quintessence potential $V(\phi)$ for the different classes. In this plot we have fixed $\beta_1 = \beta_\alpha = \beta_0 = 1$, $\gamma=0.01$ and $\delta = 1$.  \label{pot_nomatter} } }
	\includegraphics[width=0.75\linewidth]{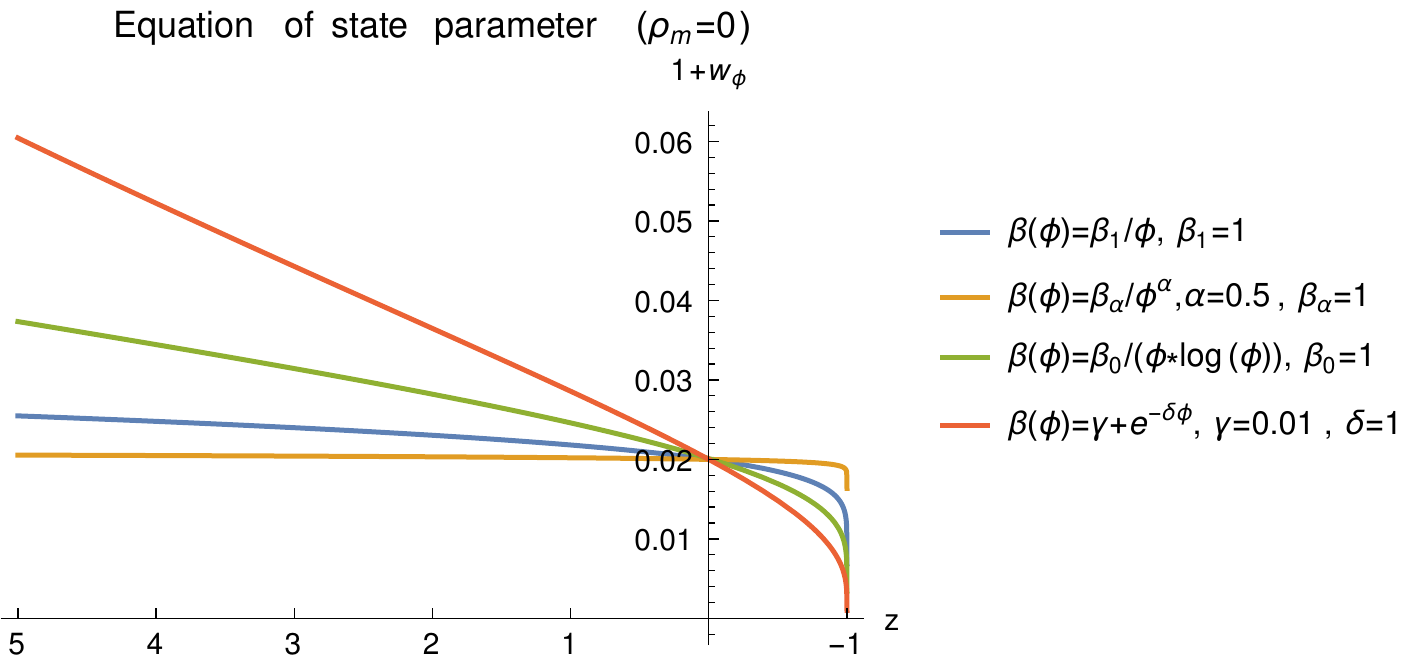}
	\caption{\footnotesize{Evolution of the equation of state parameter $w_{\phi}$. Accordingly to the choice of Fig.~\ref{pot_nomatter}, we have fixed $\beta_1 = \beta_\alpha = \beta_0 = 1$, $\gamma=0.01$ and $\delta = 1$. \label{eq_state_nomatter} }}
\end{figure}

We conclude this section by showing the asymptotic shapes of the potentials (Fig.~\ref{pot_nomatter}) and of the equation of state parameters (Fig.~\ref{eq_state_nomatter}) associated with the different classes presented in this section. As it is possible to see from Fig.~\ref{eq_state_nomatter}, for all these models we fix $\phi_0=\phi|_{\mathrm{today}}$ and we impose that at $z=0$ we are in agreement with cosmological observations\footnote{The most stringent constraint on $w_{\phi}$ set by the Planck collaboration~\cite{Ade:2015xua,Ade:2015rim} is due to the TT+TE+EE combination and reads $w < -0.97$ at $95\%$. In most of the models presented in this work we use the value $w \simeq  -0.98$.}~\cite{Ade:2015xua,Ade:2015rim} (\emph{i.e.} we set $w_{\phi0} = -0.98$ that according to Eq.~\eqref{parameter_beta_definition} also implies $\beta^2(\phi_0)\simeq 0.06$). Choosing different data sets, the value of $w_{\phi}$ today might be slightly different. However, as long as $w_{\phi}>-1$, this will not significantly change the general behavior (i.e. we will still have $w_{\phi}\to -1$ as $\phi\to\infty$) and thus it will only marginal affect the analysis.
\section{From a matter dominated to a quintessence dominated Universe. \label{sec:formalism_matter}}
We now turn our attention to the transition phase, in which the contribution of the other components of the universe must be taken into account. In the following we consider a universe which is filled by the quintessence field $\phi$ and by a second fluid \emph{i.e.} matter. The total action of the system is then:
\begin{equation}
\mathcal{S}=\int\mathrm{d}^4x\;\sqrt{-g}\left[\frac{R}{2}-\frac{1}{2}g^{\mu\nu}\partial_{\mu}\phi\partial_{\nu}\phi-V(\phi)\right]+\mathcal{S}_m \ ,
\end{equation}
where $\mathcal{S}_m$ is the action describing the matter fluid. The usual definitions of energy density and pressure for the scalar field (given in Eq.~\eqref{infldens_press}) continue to hold. The matter fluid is assumed to be dust-like matter, with energy density $\rho_m$ and no pressure ($p_m=0$). Assuming once again a flat FLRW universe (without curvature) Einstein equations~\eqref{fried1},\eqref{fried2} read:
\begin{eqnarray}
	3 H^2 &=&  \rho_m + \rho_{\phi} \ , \label{qfried1} \\
	-2 \dot{H}  &=& \rho_m+\rho_{\phi}+p_{\phi}  \label{qfried2} \ ,
\end{eqnarray}
As the two fluids are not directly coupled (\emph{i.e.} they are only interacting through gravity), the energy-momentum tensors are separately conserved. This can be seen using the continuity equations for the energy densities of the two components:
\begin{align}
\dot{\rho}_{\phi}\quad &= \quad -3H(\rho_{\phi}+p_{\phi}) \ , \label{phicont} \\
\dot{\rho}_m\quad &=\quad -3H\rho_m \ . \label{mcont}
\end{align}
As we will see in the following, Eq.~\eqref{phicont} and Eq.~\eqref{mcont} can be used in order to specify the evolution of the energy densities of the two fluids. \\

As a first step, we use Eq.~\eqref{qfried1} and Eq.~\eqref{qfried2} to express:
\begin{equation}
\label{eq:p_H_dotH}
2\dot{H}+3H^2= -p_{\phi} = -\frac{1}{2}\dot{\phi}^2+V(\phi) \ .
\end{equation}
At this point, under the assumption of a piecewise monotonic evolution for $\phi$, we can invert the dependence $\phi(t)=t(\phi)$ and we write all the quantities in terms of $\phi$. Once again we define the superpotential $W(\phi)$ as in Eq.~\eqref{superpot_def} \emph{i.e.} $W = - 2 H(\phi)$. As a consequence we have $-2 \dot{H} =\dot{W}=W_{,\phi}\dot{\phi}$, so that:
\begin{equation}
\label{hamilton_jacobi_equation}
-W_{,\phi}\dot{\phi}+\frac{3}{4}W^2=-\frac{1}{2}\dot{\phi}^2+V(\phi) \ .
\end{equation}
which can be algebraically solved for $\dot{\phi}$, showing that $\dot{\phi}$ can be expressed as a function of $\phi$ \emph{only}. We proceed by using Eq.~\eqref{infldens_press} to express the potential in terms of $\rho_{\phi}$ and $\dot{\phi}$ and using Eq.~\eqref{qfried1} to get: 
\begin{equation}
V=\rho_{\phi}-\frac{1}{2}\dot{\phi}^2=3H^2-\rho_m-\frac{\dot{\phi}^2}{2} \label{pot_in_phi} \ ,
\end{equation}
so that Eq.~\eqref{hamilton_jacobi_equation} can be expressed as:
\begin{equation}
\label{eq_for_phi_dot}
 	W_{,\phi}\dot{\phi} = \dot{\phi}^2 + \rho_m  \ .
\end{equation} 
Notice that this clearly implies that $\rho_m$ can be expressed as a function of $\phi$ \emph{only} as well.\\

We can now use the definition of $\beta$-function given in Eq.~\eqref{betafunction} \emph{i.e.}:
\begin{equation}
\beta(\phi) = \frac{\textrm{d} \phi }{\textrm{d} \ln a } = -2 \frac{\dot{\phi}}{W} . \label{betafunction_matter}
\end{equation}
to express Eq.~\eqref{eq_for_phi_dot} as:
\begin{equation}
W_{,\phi} =  -\frac{2\rho_m}{\beta W}-\frac{1}{2}\beta W \ . \label{goodone} 
\end{equation}
Finally, defining $Y(\phi)\equiv W^2(\phi)/2$, Eq.~\eqref{goodone} can be expressed as:
\begin{equation}
\beta Y_{,\phi}=-\beta^2Y-2\rho_m \ . \label{mastereq}
\end{equation}
At this point, it is also interesting to notice that using Eq.~\eqref{goodone} (or alternatively Eq.~\eqref{mastereq}) it is possible to express\footnote{Notice that $W$ is negative and thus $\sqrt{4\rho_m^2/(\beta W)^2} = - 2\rho_m/(\beta W)$. } $V(\phi)$ and $\dot{\phi}$ as:
\begin{eqnarray}
	V(\phi) &=&  \frac{3}{4} W^2 \left( 1- \frac{\beta^2}{6}\right) - \rho_m \ ,  \\
	\dot{\phi}&=&  - \frac{\beta W }{2}  \ = \ W_{,\phi} + \frac{2 \rho_m }{\beta W} \ .
\end{eqnarray}
Notice that in order to obtain this parameterization for the potential, we have both used Einstein equations~\eqref{qfried1},~\eqref{qfried2} and the continuity equations~\eqref{phicont},~\eqref{mcont}. As a consequence, this expression contains dynamical information on the evolution (for example the matter energy density $\rho_m$ is explicitly appearing).\\

\begin{figure}[htb]
	\centering
	\includegraphics[width=0.72\linewidth]{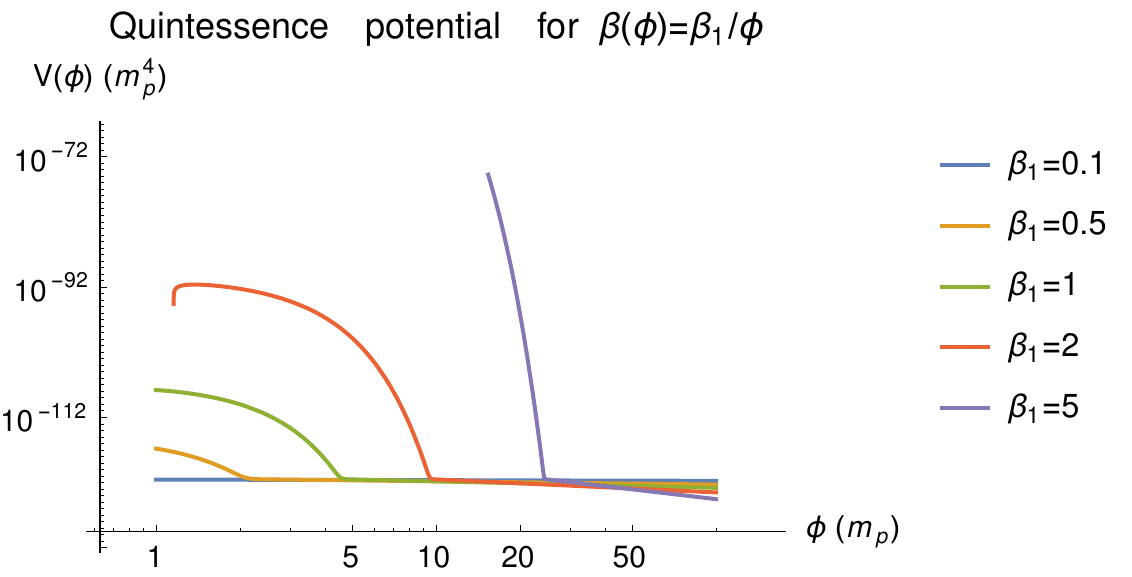}
	\caption{\footnotesize{Quintessence potential for models of the inverse monomial class considering different values of $\beta_1$.}}
	\label{fig:inverse_monomial_pot}
\end{figure}

Before proceeding with our treatment it is worth stressing two main differences with respect to the case of a Universe which is only filled by quintessence:
\begin{itemize}
	\item In general, having $\beta(\phi) \ll 1$ does not ensure the occurrence of a phase of accelerated expansion of the Universe! This can be seen by taking the ratio between Eq.~\eqref{qfried2} and Eq.~\eqref{qfried1} which gives:
\begin{equation}
	\frac{-2 \dot{H}}{3 H^2} = \frac{\rho_{\phi}+p_{\phi}}{\rho_{\phi} + \rho_m} + \frac{\rho_m}{\rho_{\phi}+ \rho_m} = \frac{\beta^2(\phi)}{3} + \frac{\rho_m/\rho_{\phi}}{1 + \rho_m/\rho_{\phi}} \ ,
\end{equation}
where we have also used the definition of $\beta$. This equation implies that a phase of accelerated expansion is realized if $\rho_m/\rho_{\phi} \ll \beta^2 \ll 1$. In the following (in particular see the discussion after Eq.~\eqref{initial}) we show that this condition is always realized for all the parameterizations of $\beta(\phi)$ which satisfy the conditions of Eq.~\eqref{eqofstate}, Eq.~\eqref{0energy} and Eq.~\eqref{infinitetime}.
\item As $\rho_m(\phi)$ is explicitly appearing in Eq.~\eqref{betafunction_matter}, the $\beta$-function (and hence the superpotential) is not only containing information on quintessence but it is also taking matter into account\footnote{In some sense this feature is similar to what happens in chameleon cosmology~\cite{Khoury:2003rn,Khoury:2013yya}, where the potential quintessence potential is combined with the coupling to matter in order to define an \emph{effective potential} $V_{\mathrm{eff}}(\phi)$.}. Therefore, by fixing a parameterization for $\beta(\phi)$ we are not simply specifying the flow of quintessence, but rather the flow of the whole system (quintessence coupled to gravity to matter). As a consequence, we only expect to reproduce the results of Sec.~\ref{sec:formalism} in the limit $\rho_m\to 0$ and elsewhere we expect significant deviations from these results. This particular feature will be distinctly visible from the profile of the potential (see Fig.~\ref{fig:inverse_monomial_pot}).
\end{itemize}

We can now compute the explicit dependence of the matter and quintessence energy densities $\rho_m$ and $\rho_{\phi}$ on $\phi$. This is done by turning the time derivatives of Eq.~\eqref{phicont} and Eq.~\eqref{mcont} into derivatives with respect to $\phi$:
\begin{eqnarray}
\dot{\rho}_m = \rho_{m,\phi}\dot{\phi}=-3H\rho_m &\implies& \frac{\rho_{m,\phi}}{\rho_m}=-3\frac{H}{\dot{\phi}}=-\frac{3}{\beta(\phi)} \ , \label{mattertophi_var}\\
\dot{\rho}_{\phi} =  \rho_{\phi,\phi}\dot{\phi}=-3H(1+w_{\phi})\rho_{\phi}&\implies& \frac{\rho_{\phi,\phi}}{\rho_{\phi}}=-\frac{3(1+w_{\phi})}{\beta(\phi)} \label{rhophibeta_var} \ ,
\end{eqnarray}
where $w_{\phi}$ is the equation of state parameter for quintessence ($p_{\phi} \equiv w_{\phi} \rho_{\phi}$) introduced in Sec.~\ref{sec:formalism}. Integrating these equations we obtain:
\begin{align}
\rho_m(\phi)\quad &= \quad\rho_{m0}\exp\left(-3\int_{\phi_0}^{\phi}\frac{\mathrm{d}\phi}{\beta(\phi)}\right) \ , \label{mattertophi} \\
\rho_{\phi}(\phi)\quad &= \quad \rho_{\phi0}\exp\left(-3\int_{\phi_0}^{\phi}\frac{1+w_{\phi}}{\beta(\phi)}\mathrm{d}\phi\right) \label{rhophibeta} \ .
\end{align}
The cosmological parameters $\Omega_m,\Omega_{\phi}$ are then defined as $\Omega_i\equiv \rho_i/(3H^2)=2\rho_i(\phi)/(3Y(\phi))$. Substituting Eq.~\eqref{mattertophi} into Eq.~\eqref{mastereq}, we get a integro-differential equation for the superpotential in terms of $\beta$. Notice that once the explicit parametrization for $\beta(\phi)$ is fixed, the evolution is completely specified. We can also express the equation of state parameter $w_{\phi}$ as a function of $\phi$. For this purpose we start from the definition of $w_{\phi}$:
\begin{equation}
1 + w_{\phi} = \frac{p_{\phi}+\rho_{\phi}}{\rho_{\phi}} = \frac{\dot{\phi}^2}{3W^2/4 - \rho_m} = \frac{\beta^2}{3} \left(1 - \frac{4 \rho_m}{3W^2} \right)^{-1}\ ,
\end{equation}
where in the last equality we used $\beta = -2 \dot{\phi}/W$. Finally we get:
\begin{equation}
1+w_{\phi}= \frac{\beta^2}{3} \left[ 1 -\frac{2\rho_{m0}}{3 Y(\phi)} \,\exp\left(\displaystyle{-3\int_{\phi_0}^{\phi}\frac{\mathrm{d}\phi'}{\beta(\phi')}}\right) \right]^{-1} \ . \label{eosparam}
\end{equation}
Again, once the $\beta$-function is specified, Eq.~\eqref{redshift_beta} can be used to compute the value of $\phi$ that correspond to a given value for $z$ so that:
\begin{equation}
1+w_{\phi}=\frac{\tilde{\beta}^2}{3} \left( 1-\frac{2\rho_{m0}}{3\tilde{Y}} \, (1+z)^3 \right)^{-1} \ ,
\end{equation}
where we have defined $\tilde{\beta}(z)\equiv\beta(\phi(z))$ and $\tilde{Y}(z)\equiv Y(\phi(z))$ \ .\\

In order to set the initial conditions for Eq.~\eqref{mastereq} we fix $\phi_0=\phi|_{\mathrm{today}}$ and we use the present values~\cite{Ade:2015xua,Ade:2015rim} $w_{\phi0}\lesssim -0.98$ and $\Omega_{m0}\simeq 0.3175$ so that:
\begin{align}
Y_0\quad &=\quad 2H_0^2,\qquad \qquad H_0\simeq 1.18\times 10^{-61}\; (\mbox{in Planck units}) \notag \\
\rho_{m0}\quad &=\quad 3H_0^2\Omega_{m,0}=0.9525 H_0^2 \ , \notag \\
\left.1+w_{\phi}\right|_{\phi_0}\quad &=\quad \frac{\beta^2(\phi_0)/3}{1 -2\rho_{m0}/(3Y_0)}=\frac{\beta^2(\phi_0)/3}{1-\rho_{m0}/(3H_0^2)}=0.02 \ ,
\end{align}
where we have defined $Y_0 \equiv Y(\phi_0)$. Hence we get:
\begin{equation}
\beta^2(\phi_0) \simeq 0.04095 \ . \label{initial}
\end{equation}
As $\rho_{m0} < \rho_{\phi0} $ and $w_{\phi0} < 0 $, Eq.~\eqref{mattertophi_var} and Eq.~\eqref{rhophibeta_var} imply that $\rho_m$ decrease faster than $\rho_\phi$. As a consequence for $\phi > \phi_0$ we can safely approximate $1 + w_\phi \simeq \beta^2 / 3$ and (taking the ratio between Eq.~\eqref{mattertophi} and Eq.~\eqref{rhophibeta}) the condition $\rho_m/\rho_\phi \ll \beta^2 \ll 1$ simply reads:
\begin{equation}
	\label{mattertophiratio}
	\frac{\rho_m(\phi)}{\rho_{\phi}(\phi)} \simeq \frac{\rho_{m0}}{\rho_{\phi0}} \exp\left(-3\int_{\phi_0}^{\phi}\frac{ \mathrm{d}\phi}{\beta(\phi)}\right) \ll \beta^2(\phi) \ll 1 \ ,
\end{equation}
which is obviously satisfied by all the parameterizations of $\beta(\phi)$ which satisfy the conditions of Eq.~\eqref{eqofstate} Eq.~\eqref{0energy} and Eq.~\eqref{infinitetime}. In particular it is satisfied by all the parameterizations\footnote{Moreover, as matter naturally provides a mechanism to exit from the phase of accelerated expansion, a strictly constant $\beta$-function is allowed in this framework! The predictions for this case (referred to as power-law class, in contrast with the case presented in Sec.~\ref{sec:univ-quint}) are shown in Appendix~\ref{sec:appendix_other} (in particular see Sec.~\ref{sec:appendix_power_law}).} introduced in Sec.~\ref{sec:univ-quint}.\\

\begin{figure}[htb]
	\centering
	\includegraphics[width=1.3\linewidth]{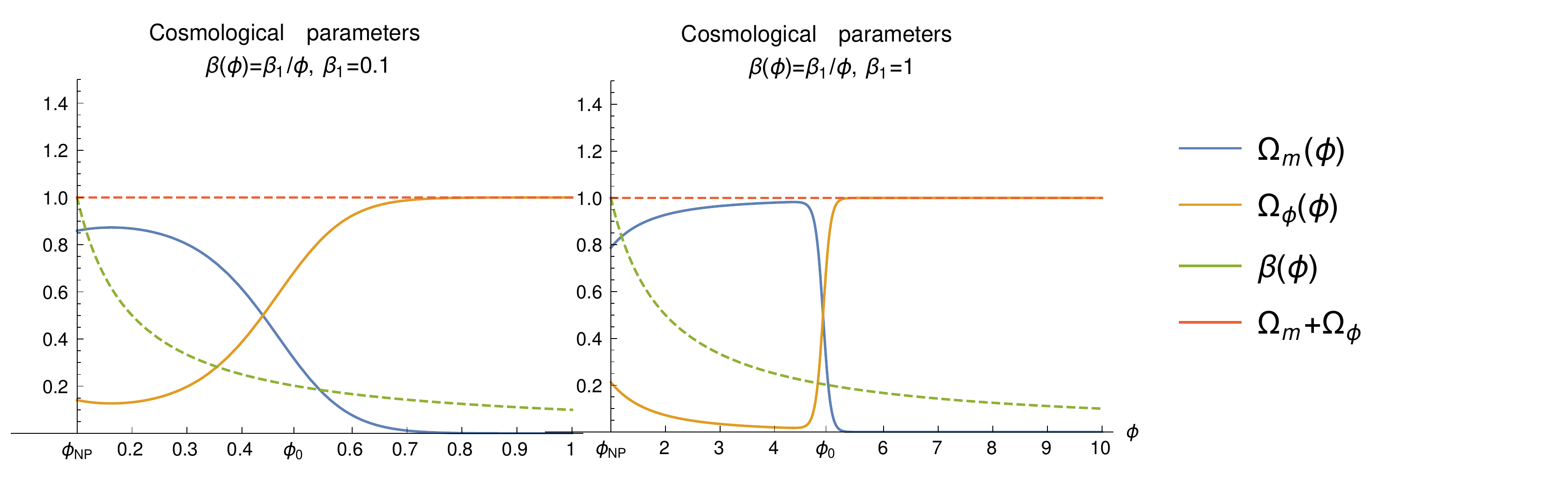}
	\caption{\footnotesize{Evolution of the cosmological parameters for the inverse monomial class with $\beta_1=0.1$ (left) and $\beta_1=1$ (right).}}
	\label{fig:comparison_monomial}
\end{figure}

\begin{figure}[htb]
	\centering
	\includegraphics[width=0.65\linewidth]{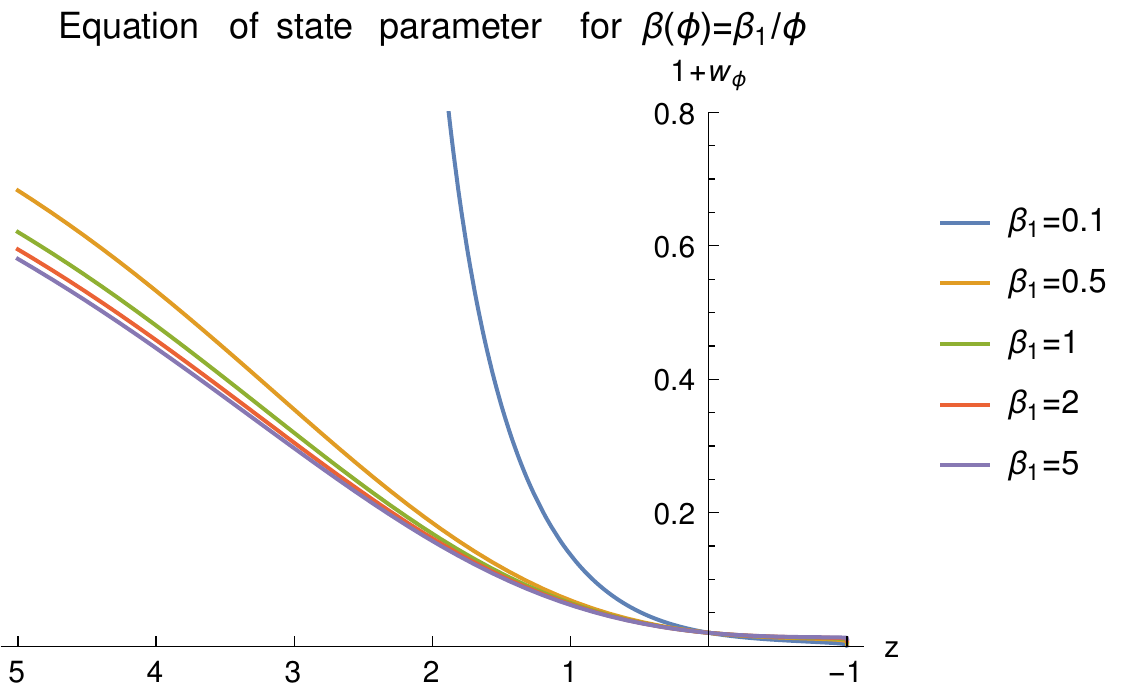}
	\caption{\footnotesize{Evolution of the equation of state parameter for the models of the inverse monomial class considering different values of $\beta_1$.}}
	\label{fig:inverse_monomial_eq_of_state}
\end{figure}

To conclude this section we focus our discussion on one of the classes presented in Sec.~\ref{sec:formalism}, namely the inverse monomial class (the analysis for the other classes is presented in Appendix~\ref{sec:appendix_other}) for which the expansion of the $\beta$-function near the fixed point is given by $\beta(\phi)\simeq\beta_1/\phi$. In the plot of Fig.~\ref{fig:inverse_monomial_pot} we show the shape of the potential for some values of the parameter $\beta_1$. In particular it is possible to notice that for models with $\beta_1>\mathcal{O}(10^{-1})$, the potential exhibits a transition (at $\phi \simeq \phi_0$) that corresponds to the transition from the matter-dominated epoch to the quintessence-dominated epoch. During the latter, the effects induced by matter are negligible and $V(\phi)$ reduces to the inverse monomial potential shown in Eq.~\eqref{inverse_monom_pot}. Whether this transition is present or not has important consequences on the evolution of the cosmological parameters $\Omega_m,\Omega_{\phi}$. In Fig.~\ref{fig:comparison_monomial}, we show the evolution of $\Omega_m$ and $\Omega_{\phi}$ for two different values of $\beta_1$. In the case with $\beta_1=0.1$, we are not able to see a phase of matter dominance (\emph{i.e.} $\Omega_m \simeq 1$ and $\Omega_{\phi} \simeq 0$) before having $\beta(\phi) \simeq 1$, where the asymptotic expansion for $\beta(\phi)$ ceases to be valid. Notice that this model clearly exhibits a transition from matter to quintessence domination at $\phi \simeq \phi_0$ for $\beta_1 > 0.1$. Finally in Fig.~\ref{fig:inverse_monomial_eq_of_state} we show the evolution of the equation of state parameter as a function of redshift $z$ for the models of Fig.~\ref{fig:inverse_monomial_pot}. Notice that in order to respect cosmological observations, all models meet at $1+w_{\phi} = 0.02$ at present time ($z=0$).

\section{Non-minimal coupling. \label{sec:formalism_non_min}}
In this Section we discuss the possibility of having a non-minimal coupling $\Omega(\phi)$ between quintessence and gravity. Let us assume that the action in the Jordan frame is:
\begin{equation}
\mathcal{S}_J=\int\mathrm{d}^4x\;\sqrt{-\tilde{g}}\left[\frac{\Omega(\phi)}{2}\tilde{R}-\frac{1}{2}\tilde{g}^{\mu\nu}\partial_{\mu}\phi\partial_{\nu}\phi-V_J(\phi)\right] + \tilde{\mathcal{S}}_{mJ} \ , \label{jordan}
\end{equation}
where $\tilde{\mathcal{S}}_{mJ}$ denotes the action for matter in the Jordan frame. We assume that the matter action in the Jordan frame $\tilde{\mathcal{S}}_{mJ}$ does not depend on $\phi$. We can start by performing a conformal transformation of the metric:
\begin{equation}
\tilde{g}_{\mu\nu}\to g_{\mu\nu}=\Omega(\phi)\tilde{g}_{\mu\nu} \ , \label{conformal}
\end{equation}
to bring the action of Eq.~\eqref{jordan} in the Einstein frame, where gravity is described by a standard Einstein-Hilbert term:
\begin{align}
\mathcal{S}_E=&\int\mathrm{d}^4x\;\sqrt{-g}\left[\frac{R}{2}-\left(\Omega^{-1}+\frac{3}{2}\left(\dev{\ln \Omega}{\phi}\right)^2\right)\frac{1}{2}g^{\mu\nu}\partial_{\mu}\phi\partial_{\nu}\phi- V(\phi) \right] + \tilde{\mathcal{S}}_{mE}  \ , \label{einsteinf}
\end{align}
where $ V(\phi) \equiv V_J(\phi) / \Omega^2(\phi) $. Notice that as the metric was redefined according to Eq.~\eqref{conformal}, the action for matter in the Einstein frame depends on $\phi$! As a consequence matter and quintessence are now coupled and thus the energy-momentum tensors are not independently conserved. \\

At this point it is useful to define a new field $\varphi$ as:
\begin{equation}
\left(\dev{\varphi}{\phi}\right)^2=\Omega^{-1}+\frac{3}{2}\left(\dev{\ln \Omega}{\phi}\right)^2 \ , 
\end{equation}
so that kinetic term for quintessence is cast in the canonical form:
\begin{equation}
\mathcal{S}_E=\int \mathrm{d}^4x\;\sqrt{-g}\left[\frac{R}{2}-\frac{1}{2}g^{\mu\nu}\partial_{\mu}\varphi\partial_{\nu}\varphi- \bar{V}(\varphi) \right]  + \mathcal{S}_{mE}  \ ,  \label{einsteincan}
\end{equation}
where we have defined $ \bar{V}(\varphi) \equiv V(\phi(\varphi))$ and the matter action $\mathcal{S}_{mE}$ is now depending explicitly on $\varphi$. \\

Assuming that the Einstein frame metric $g_{\mu\nu}$ is a flat FLRW\footnote{Although we are not interested in describing the system in the Jordan frame, we should note that (as discussed in~\cite{EspositoFarese:2000ij}) through a redefinition of the time coordinate and of the scale factor, it is possible to work with a FLRW metric both in the Jordan and Einstein frame. More details on this point are given in Appendix~\ref{esposito}.}
\begin{equation}
\mathrm{d}s_E^2=-\mathrm{d}t^2+a^2(t)\mathrm{d}\mathbf{x}^2 \ ,
\end{equation}
that both quintessence and matter are homogeneous and that matter is dust-like ($p_m=0$) we get: 
\begin{eqnarray}
	3 H^2 &=&  \rho_m + \rho_{\varphi} \ , \label{friedNM} \\
	-2 \dot{H}  &=& \rho_m+\rho_{\varphi}+p_{\varphi}  \label{einstein_2} \ ,
\end{eqnarray} 
where $p_{\varphi} $ and $\rho_{\varphi}$ are respectively the pressure and the energy densities associated with $\varphi$ \emph{i.e.}:
\begin{equation}
\rho_{\varphi}  =  \frac{\dot{\varphi}^2}{2}+\bar{V}(\varphi) \ , \qquad  \qquad p_{\varphi}  =  \frac{\dot{\varphi}^2}{2}-\bar{V}(\varphi) \ , \label{phi_energy_press} 
\end{equation}
and $\rho_m$ is the matter energy density. At this point we can use Eq.~\eqref{friedNM}, Eq.~\eqref{einstein_2} and Eq.~\eqref{phi_energy_press} to express:
\begin{equation}
	3H^2 + 2\dot{H} = -p_{\varphi} = - \frac{\dot{\varphi}^2}{2} + V \ ,
\end{equation}
As this equation is equal to the one obtained in the case of minimal coupling (Eq.~\eqref{eq:p_H_dotH}) we can directly follow the treatment of Sec.~\ref{sec:formalism_matter}. In particular, after the introduction of the superpotential, the expressions for $\dot{\varphi},\bar{V}$ and $\beta(\varphi)$ are the same as in the minimally coupled case. Moreover, the equation that relates $W$ and $\beta$ simply reads:
\begin{equation}
W_{,\varphi}=-\frac{2\rho_m}{\beta W}-\frac{1}{2}\beta W, \quad \iff \quad \beta Y_{,\varphi}= -2\rho_m -\beta^2Y \ .
\end{equation}
Notice that this equation has exactly the same form of Eq.~\eqref{mastereq}. However, it is crucial to stress that as in this case the two fluids are coupled, the evolution of $\rho_{m}$ and $\rho_{\varphi}$ (which is set by the continuity equations) is different! In particular, in order to fix the evolution of the two energy species we need to specify the characteristics of the matter action $\tilde{\mathcal{S}}_{mJ}$.\\

In the following we focus on the case in which matter is described by a homogeneous field (assumed to be scalar). In particular we assume that in the Jordan frame the action $\tilde{\mathcal{S}}_{mJ}$ is given by:
\begin{equation}
\tilde{\mathcal{S}}_{mJ}=\int\mathrm{d}^4x\;\sqrt{-\tilde{g}}\left[ -\frac{1}{2}\tilde{g}^{\mu\nu}\partial_{\mu}\psi\partial_{\nu}\psi-U(\psi) \right]  \ , \label{matter_jordan}
\end{equation}
As a consequence it is easy to show that 
\begin{align}
\tilde{\mathcal{S}}_{mE}=&\int\mathrm{d}^4x\;\sqrt{-g}\left[-\frac{g^{\mu\nu}\partial_{\mu}\psi\partial_{\nu}\psi}{2\Omega(\phi)}-\frac{U(\psi)}{\Omega^2(\phi)}\right] \ , \label{einsteinf_matter_phi} \\
\mathcal{S}_{mE} =& \int \mathrm{d}^4x\;\sqrt{-g}\left[-\frac{g^{\mu\nu}\partial_{\mu}\psi\partial_{\nu}\psi}{2 \Psi(\varphi)}-\frac{U(\psi)}{\Psi^2(\varphi)}\right] \ ,\label{einsteinf_matter_varphi}
\end{align}
where we have defined $\Psi(\varphi) \equiv \Omega(\phi(\varphi))$. At this point we can express the energy density and pressure associated with $\psi$ in the Einstein frame:
 \begin{eqnarray}
\rho_m &=& \frac{\dot{\psi}^2}{2\Psi(\varphi)}+\frac{U(\psi)}{\Psi^2(\varphi)} \ , \label{mattdensNM} \\
p_m &=& \frac{\dot{\psi}^2}{2\Psi(\varphi)}-\frac{U(\psi)}{\Psi^2(\varphi)} \ . \label{mattpressNM} 
\end{eqnarray}
As already explained in this section, since the fluids are coupled, the energy-momentum tensors are not conserved independently. As a consequence, in order to get the continuity equations, we should start by computing the equations of motion for the two scalar fields:
 \begin{eqnarray}
\ddot{\varphi} &=& -3H\dot{\varphi}-\bar{V}_{,\varphi}(\varphi)-\frac{\Psi_{,\varphi}}{2\Psi^2}\dot{\psi}^2+\frac{2\Psi_{,\varphi}}{\Psi^3}U(\psi) \label{kleinphi}, \\
\ddot{\psi} &=& -3H\dot{\psi}+\frac{\dot{\Psi}}{\Psi}\dot{\psi}-\frac{U_{,\psi}(\psi)}{\Psi}. \label{kleinpsi}
\end{eqnarray}
Using the assumption of dust-like matter ($p_m = 0$) we easily get:
\begin{equation}
	\label{mattdens_dust}
 	\rho_m = \frac {2 U(\psi)}{\Psi^2} = \frac{\dot{\psi}^2}{\Psi(\varphi)} \ ,
 \end{equation} 
so that differentiating the energy densities for $\varphi$ and $\psi$ (given by Eq.~\eqref{phi_energy_press} and Eq.~\eqref{mattdens_dust}) with respect to time, we obtain the continuity equations:
\begin{eqnarray}
\dot{\rho}_{\varphi} &=& -3H(\rho_{\varphi}+p_{\varphi})+\frac{1}{2}\dev{\ln \Psi(\varphi)}{t}\rho_m \ , \label{phicontNM} \\
\dot{\rho}_m &=& -\left(3H+\frac{1}{2}\dev{\ln \Psi(\varphi)}{t}\right)\rho_m \ . \label{mattcontNM}
\end{eqnarray}
As expected the two fluids are coupled and in particular this coupling is proportional to the factor $(\mathrm{d\ln \Psi}/\mathrm{d}t)\rho_m/2$. \\

It is crucial to stress that in this case the dependence of $\rho_m$ on $\varphi$ is different from the case of minimal coupling! In fact we have:
\begin{equation}
\label{rho_m_non_minimal}
\rho_m(\varphi)\quad =\quad \left(\frac{\Psi_0}{\Psi(\varphi)}\right)^{1/2}\rho_{m0}\exp\left(-3\int_{\varphi_0}^{\varphi}\frac{\mathrm{d}\varphi'}{\beta(\varphi')}\right) \ .
\end{equation}
The equation-of-state parameter $w_{\varphi}$ is thus given by:
\begin{equation}
1+w_{\varphi}= \beta^2(\varphi)Y(\varphi) \left[ 3Y(\varphi)-2\rho_{m0}\left(\dfrac{\Psi_0}{\Psi(\varphi)}\right)^{1/2}\exp\left(\displaystyle{-3\int_{\varphi_0}^{\varphi}\frac{\mathrm{d}\varphi'}{\beta(\varphi')}}\right) \right]^{-1} \ .
\end{equation}
While the initial conditions (which are set at $\varphi_0=\varphi_{\mathrm{today}}$) remain unchanged:
\begin{align}
Y(\varphi_0)&=2H_0^2, \notag \\
\rho_{m0}&\simeq 0.9525 H_0^2, \notag \\
\beta^2(\varphi_0)&=0.04095 \ ,
\end{align}
in order to completely specify the evolution we need to fix an explicit parametrization for $\beta(\varphi)$ and for the non-minimal coupling $\Psi(\varphi)$.

\subsection{Attractors at strong coupling.}
\label{sec:strong_attractors}
Interestingly, in some particular cases even if a non-minimal coupling is present, the evolution is still completely specified by $\beta(\varphi)$. For example such a scheme is realized in presence of an attractor. In the context of inflation several models that give rise to attractors are known. In particular it is worth mentioning the class of models defined by Kallosh and Linde in~\cite{Kallosh:2013xya,Kallosh:2013hoa,Kallosh:2013daa} and subsequently generalized by Kallosh, Linde and Roest in~\cite{Kallosh:2013yoa,Kallosh:2014rga}, which lead to the appearance of the well known ``$\alpha$-attractors''. Another interesting class of models has been recently proposed by Linde, Kallosh and Roest~\cite{Kallosh:2013tua} and it leads to the appearance of the so-called attractor at strong coupling. While the analysis of these models in terms of the $\beta$-function formalism for inflation was carried out in~\cite{Pieroni:2015cma}, in the following we show that a similar parametrization can be used to define a class of models for quintessence. \\

Let us start by expressing the non-minimal coupling $\Omega(\phi)$ as~\cite{Kallosh:2013tua}:
\begin{equation}
\Omega(\phi)=1+\xi f(\phi) \ , \qquad \Psi(\varphi)=1+\xi f(\phi(\varphi))\equiv 1+\xi\bar{f}(\varphi) \ .
\end{equation}
where $f(\phi)$ is a general function of $\phi$. In this way, we have
\begin{equation}
\label{non_minimal_factor}
\left(\dev{\varphi}{\phi}\right)^2=F(\phi)=\frac{1+\xi f(\phi)+3\xi^2f^2_{,\phi}(\phi)/2}{(1+\xi f(\phi))^2}.
\end{equation}
In the limit of $\xi f(\phi) \gg 1$, Eq.~\eqref{non_minimal_factor} can be approximated~\cite{Kallosh:2013tua,Pieroni:2015cma} as:
\begin{equation}
\left(\dev{\varphi}{\phi}\right)^2=F(\phi)=\Omega^{-1}(\phi)+\frac{3}{2}\left(\dev{\ln \Omega}{\phi}\right)^2 \simeq \frac{3}{2}\left(\frac{\Omega_{,\phi}}{\Omega}\right)^2 \simeq \frac{3}{2}\left(\frac{f_{,\phi}}{f}\right)^2 \ ,
\end{equation}
by taking the square root of this equation\footnote{This equation has two solutions, which correspond to whether the fixed point is reached at $+\infty$ or $-\infty$. Assuming that the fixed point is reached at $\phi\to+\infty$, we should set: 
	\begin{equation}
\sqrt{\frac{3}{2}}\frac{f_{,\phi}}{f}=-\dev{\varphi}{\phi} \ .
\end{equation}} 
and integrating we directly get:
\begin{equation}
\label{f_parametrization_strong}
\bar{f}(\varphi) \equiv f(\phi(\varphi))=\bar{f}_0\exp\left[-\sqrt{2/3}(\varphi-\varphi_0)\right] \ .
\end{equation}
Remarkably, for $\xi \gg 1$ the expression of $\bar{f}(\varphi)$ does not depend on the particular choice of $f(\phi)$. As a consequence we do not need to specify the parameterization of $\Psi(\varphi)$ and the evolution is completely specified by the $\beta$-function! For example in the case of the attractor at strong coupling~\cite{Kallosh:2013tua}, by fixing the Jordan frame potential to be\footnote{This parametrization for $V_J(\phi)$ is motivated by the possibility of defining a natural embedding of these models in supergravity~\cite{Kallosh:2010ug}.}:
\begin{equation}
  	V_J(\phi) = \lambda^2 f^2(\phi) \ ,
  \end{equation}  
it is possible to show~\cite{Pieroni:2015cma} that independently on the explicit parametrization for $f(\phi)$ we have:
\begin{equation}
	\bar{\beta}(\varphi) = \beta(\phi(\varphi)) = \exp\left[-\sqrt{2/3}(\varphi-\varphi_0)\right] \ .
\end{equation}
While this model is extremely interesting in the context of inflationary model building, in order to describe models of quintessence we have to consider different parameterizations for $\beta(\varphi)$. In particular we have to consider parameterizations of $\beta(\varphi)$ that respect the conditions set by Eq.~\eqref{eqofstate}, Eq.~\eqref{0energy} and Eq.~\eqref{infinitetime} (such as the classes introduced in Sec.~\ref{sec:formalism}). Before specifying an explicit expression for $\beta(\varphi)$ we can compute (substituting Eq.~\eqref{f_parametrization_strong} into Eq.~\eqref{rho_m_non_minimal}):
\begin{equation}
\rho_m(\varphi)= \rho_{m0}\exp\left(\frac{\varphi-\varphi_0}{\sqrt{6}}-3\int_{\varphi_0}^{\varphi}\frac{\mathrm{d}\varphi'}{\beta(\varphi')}\right) \ ,
\end{equation}
and, for completeness, we also give the expression for $Y = W^2/2$: 
\begin{align}
Y(\varphi)&=2\exp\left(-\int_{\varphi_0}^{\varphi}\beta(\varphi')\mathrm{d}\varphi'\right)\left\{H_0^2-\rho_{m0}\int_{\varphi_0}^{\varphi}\frac{\mathrm{d}\varphi'}{\beta(\varphi')}\times \right. \notag \\
&\left. \exp\left[\frac{\varphi'-\varphi_0}{\sqrt{6}}+\int_{\varphi_0}^{\varphi'}\left(-\frac{3}{\beta(\varphi'')}+\beta(\varphi'')\right)\mathrm{d}\varphi''\right]\right\},
\end{align}
where we used the initial condition $Y_0=2H_0^2$.\\

To conclude this section we consider an explicit parametrization for the $\beta$-function (in terms of $\varphi$) and we show the predictions for the quintessence potential, the equation of state parameter and for the evolution of the energy densities of matter and quintessence. Again we consider the case of the inverse monomial class\footnote{The results for the other classes introduced in Sec.~\ref{sec:formalism} are not shown in this work. However, similarly to the case of the inverse monomial class, the introduction of a non-minimal coupling between quintessence and gravity does not significantly affect the results shown in Appendix~\ref{sec:appendix_other}.} introduced in Sec.~\ref{sec:formalism} for which the expansion of the $\beta$-function near the fixed point is given by $\beta(\varphi)\simeq\beta_1/\varphi$.\\

\begin{figure}[h]
	\centering
	\includegraphics[width=0.7\linewidth]{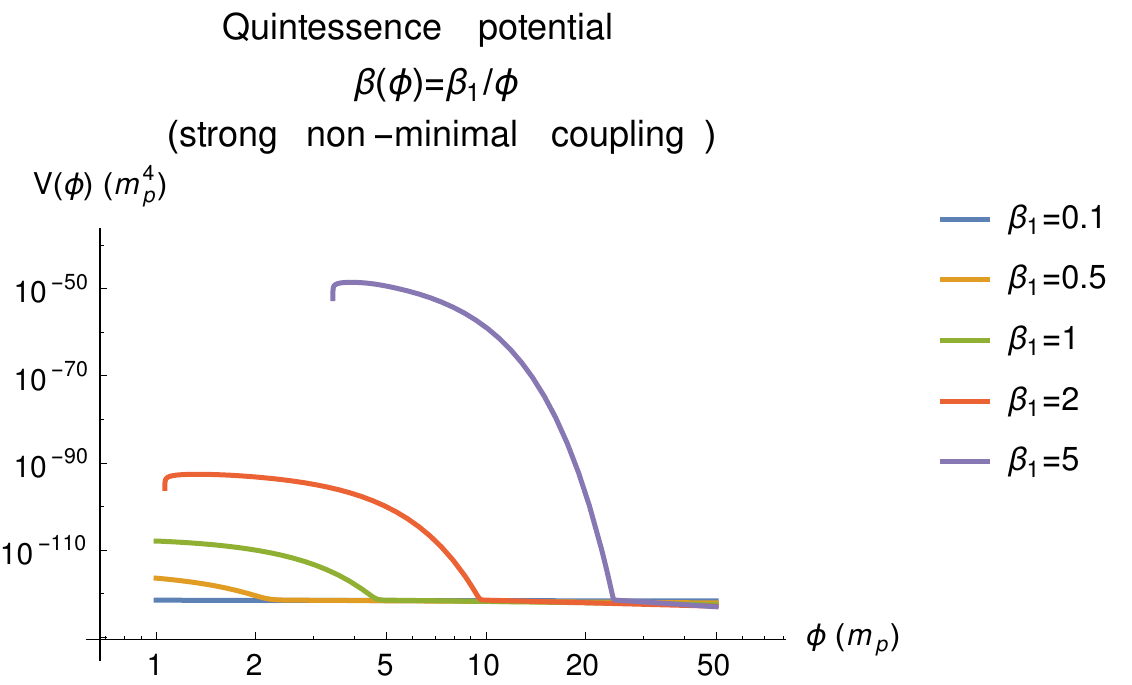}
	\caption{\footnotesize{Profile of the quintessence potential for models of the inverse monomial class considering different values of $\beta_1$.}}
	\label{fig:potential_non_minimal}
\end{figure}

\begin{figure}[h]
	\centering
	\includegraphics[width=0.68\linewidth]{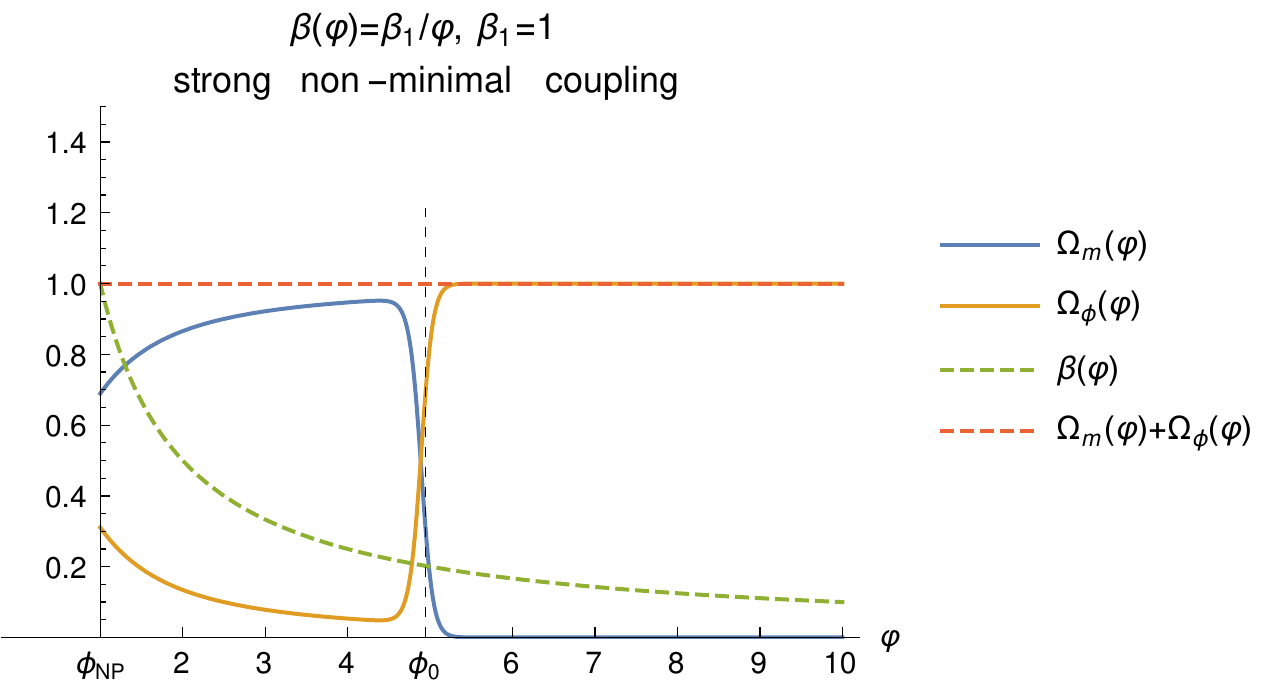}
	\caption{\footnotesize{Evolution of the energy densities for $\beta_1 = 1$.}}
	\label{fig:non_minimal_evolution}
\end{figure}

\begin{figure}[h]
	\centering
	\includegraphics[width=0.44\linewidth]{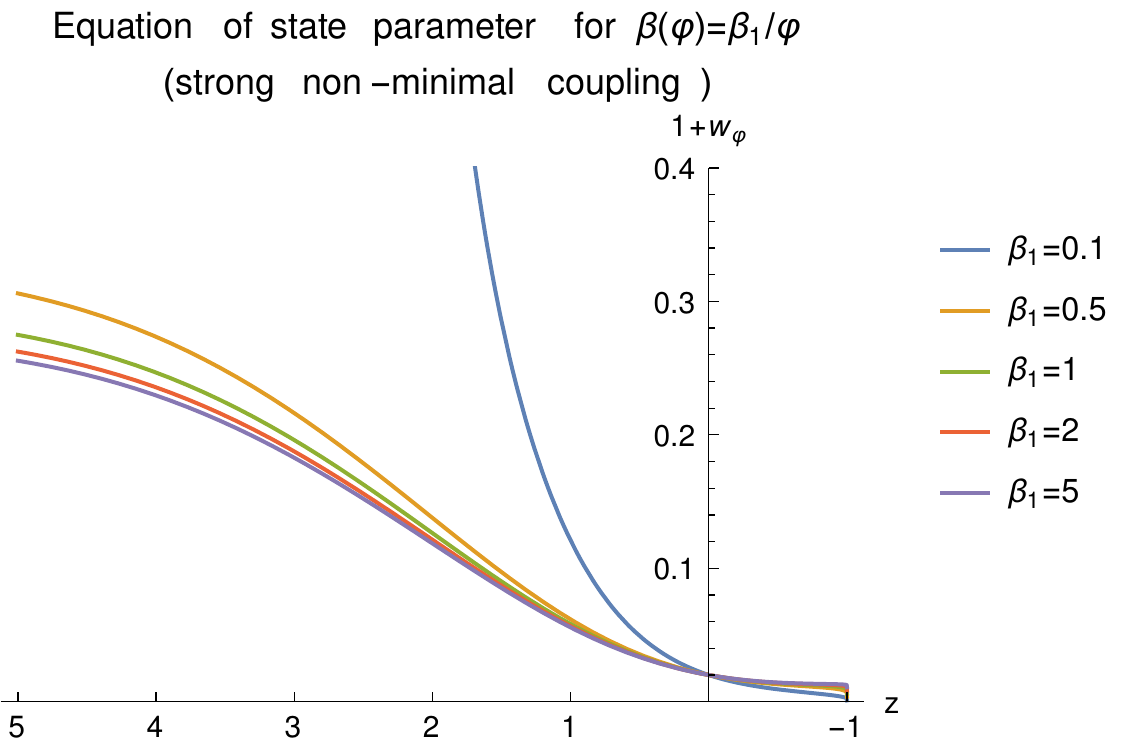}
	\includegraphics[width=0.55\linewidth]{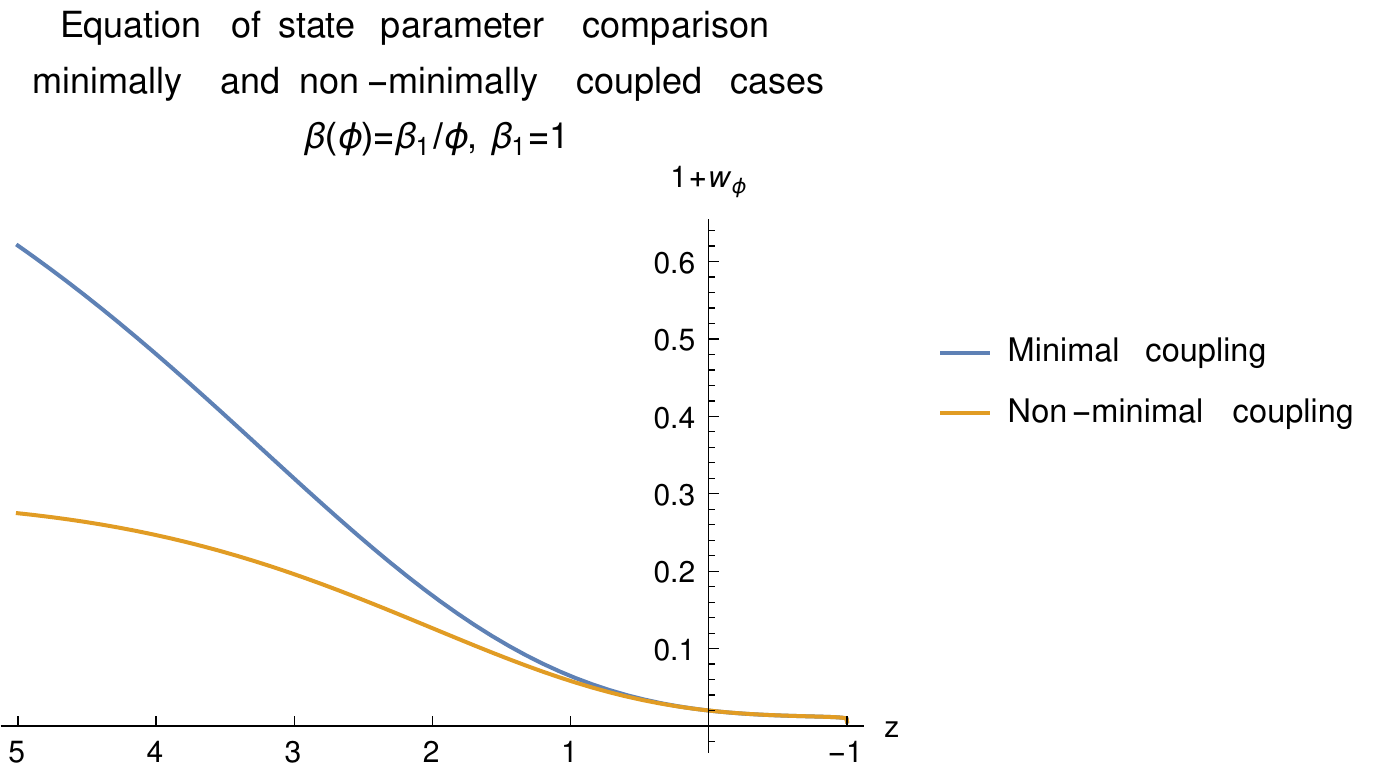}
	\caption{\footnotesize{Evolution of the equation of state parameter for some models of the inverse monomial class with non-minimal coupling (on the left) and comparison between a minimally coupled and a non-minimally coupled model with $\beta_1 =1$ (on the right).}}
	\label{fig:eq_of_state_non_minimal}
\end{figure}

As it is possible to see from Fig.~\ref{fig:potential_non_minimal}, for $\beta_1 > 0.1$ the quintessence potential shows again a transition (similar to the one shown in Fig.~\ref{fig:inverse_monomial_pot}). Notice however, that in the matter dominated phase the value of $V(\phi)$ is larger than in the minimally coupled case (see Fig.~\ref{fig:inverse_monomial_pot}). Indeed, this feature affects the evolution of the cosmological parameters\footnote{In this plot we are only presenting the evolution for the case $\beta_1 =1$. The other models show a similar behavior except for the case with $\beta_1 =0.1$ where the matter-dominated phase is not reproduced.} shown in Fig.~\ref{fig:non_minimal_evolution}. As expected we have quintessence dominance for $\varphi>\varphi_0$ and matter dominated epoch is correctly reached for $\varphi_{\mathrm{NP}}<\varphi<\varphi_0$. However, the maximum value of $\Omega_m$ in this case is slightly smaller than in the minimally coupled case (see Fig.~\ref{fig:comparison_monomial}). This is probably due to the larger value of the quintessence potential in this regime. Finally, in Fig.~\ref{fig:eq_of_state_non_minimal}, we show the evolution of the equation-of-state parameter versus the redshift. Generally, the value of $1+w_{\phi}$ is smoother than in the minimally coupled case\footnote{This suggest that the evolution of a non-minimally coupled quintessence fluid is smoother. As a consequence this may give a further hint to explains why $\Omega_m$ does not reach the value of 1 in Fig.~\ref{fig:non_minimal_evolution} before the breaking of the perturbative regime.} (showed in Fig.~\ref{fig:inverse_monomial_eq_of_state}). Nevertheless, all the curves correctly tend to zero in the infinite future.
\section{Conclusions.\label{sec:conclusions}}
During the next decade, with the new generation of galaxy surveys (such as Euclid and LSST) we expect to be able to accurately measure the acceleration of the Universe. These observations offer an extremely powerful method to probe modifications of general relativity on cosmological scales and thus they may help us to shed some light on the nature of DE. As a consequence it is crucial to study all the possible landscapes and moreover it is useful to define a method to set a direct connection between theory and observations. \\

The $\beta$-function formalism provides a powerful tool for studying various cosmological landscapes. In this work we have discussed its application to the case of quintessence. After the definition of a minimal set of universality classes for models of quintessence (in Sec.~\ref{sec:univ-quint}), we have studied the cosmological evolution of a Universe that is both filled by matter and quintessence. In particular we have shown (in Sec.~\ref{sec:formalism_matter}) that once the initial conditions (given by present time observations) are set, the transition from a phase of matter domination to a phase dominated by quintessence naturally occurs in most of the classes. Moreover, we have shown (in Sec.~\ref{sec:formalism_non_min}) that this behavior does not change significantly if quintessence is non-minimally coupled to gravity. Interestingly, we find the evolution of the equation of state parameter appears to be smoother if quintessence is non-minimally coupled to gravity. This feature can help in distinguishing between these two scenarios.\\

In terms of the $\beta$-function formalism we have a powerful bottom-up method to construct new models for quintessence. We have shown this framework can be naturally extended in order to describe more realistic situations where other components (such as matter) affect the evolution of the universe. Moreover, as $\beta$ is not only controlling the evolution of quintessence but rather the evolution of the whole system, it defines a particularly convenient framework to describe the late time evolution of our Universe. \\

The discussion presented in this paper was carried out following a phenomenological approach. However, theoretical argument to support the introduction of this formalism arise if we consider an holographic description of cosmology. In particular, following the procedure described in~\cite{McFadden:2009fg,McFadden:2010na,Pieroni:2016gdg}, the AdS/CFT correspondence of Maldacena~\cite{Maldacena:1997re} can be applied to cosmology and the approaching of a (A)dS configuration (which in the case of quintessence is realized in the infinite future) corresponds to the approaching of conformal invariance for the dual (pseudo-)QTF. In this framework it is possible to carry out an holographic analysis of the RG flow induced by the different universality classes. This discussion may help to shed some light on the possibility of defining a consistent embedding of these models in some high energy theory.\\

While the generalization of the formalism to models with non-standard kinetic terms is carried out in~\cite{Binetruy:2016hna}, generalizations to multi-fields models\footnote{The analyses of~\cite{Bourdier:2013axa} and of~\cite{Garriga:2015tea} can provide useful guides to achieve this result.}, to models with a non-minimally \emph{kinetic} coupling between the quintessence and gravity (as in the case of Horndeski theories~\cite{Horndeski:1974wa,Deffayet:2011gz,Kobayashi:2011nu}), and to models of modified gravity are still lacking. These possibility offer interesting prospects for future works.

\subsection*{ Acknowledgements}
We would like to thank Pierre Bin\'etruy for useful and stimulating discussions which provived an indispensable guide for the development of this work. M.P. wants to thank Valerie Domcke for her instructive observations and F.C. wants to thank Andrea Ferrara for his useful suggestions. M.P. acknowledges the financial support of the UnivEarthS Labex program at Sorbonne Paris Cit\'e (ANR-10-LABX-0023 and ANR-11-IDEX-0005-02). 

\newpage
\appendix

\section{Other cases. \label{sec:appendix_other}}
In this appendix, we discuss some other parametrization of the $\beta$-function among those presented in Sec.~\ref{sec:univ-quint}. The plots shown in this section only refer to the minimally coupled case. However, it is possible to show that the introduction of a non-minimal coupling between quintessence and gravity is not significantly affecting these results.

\subsection{Inverse fractional class.}
\begin{figure}[h]
\centering
\includegraphics[width=0.495\linewidth]{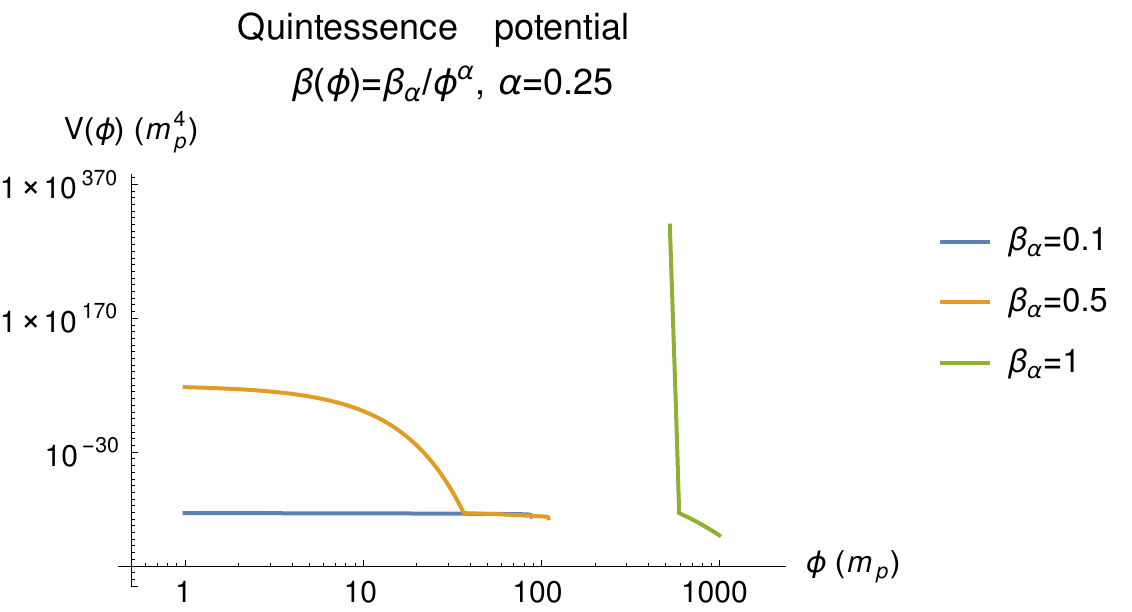}
\includegraphics[width=0.495\linewidth]{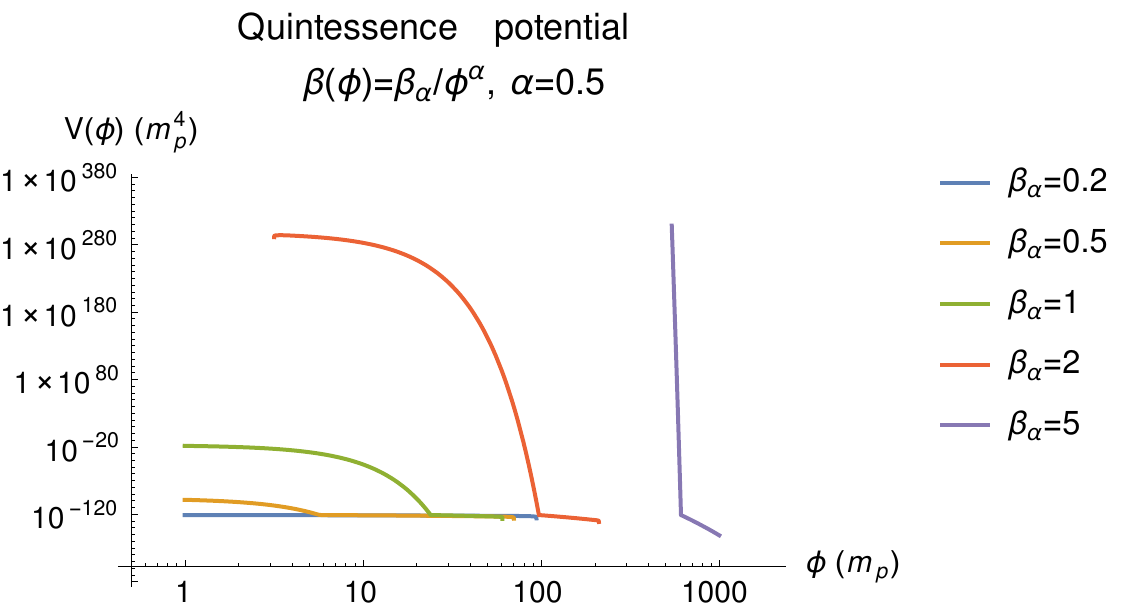}
\includegraphics[width=0.495\linewidth]{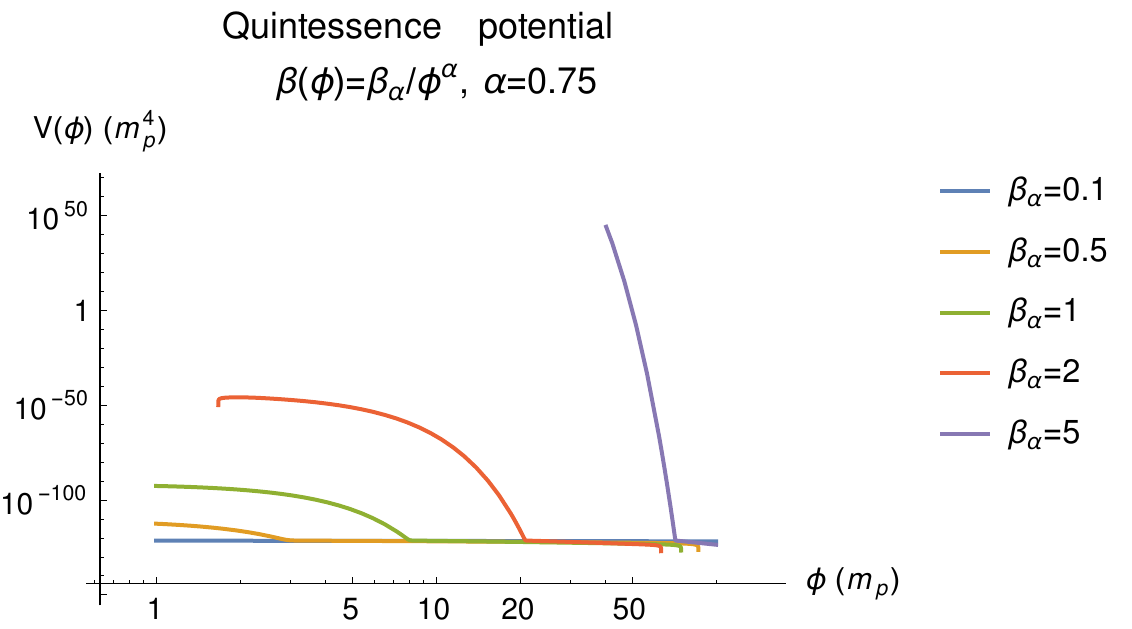}
\caption{Evolution of the quintessence potential in the inverse fractional class for different choices of $\alpha$ and $\beta_{\alpha}$.}
\label{fig:potentials_fractional}
\end{figure}

The expansion of the $\beta$-function for this class reads $\beta(\phi)=\beta_{\alpha}/\phi^{\alpha} + o(1/\phi^{\alpha})$ with $0<\alpha<1$. In Fig.~\ref{fig:potentials_fractional} we show the shape of the quintessence potential for different values of $\alpha$ and $\beta_{\alpha}$. While the value of $\alpha$ does not significantly affects the shape of $V(\phi)$, it strongly affects its amplitude. Moreover, we see that in general for $\beta_{\alpha}>0.2$ the matter-quintessence transition in $V(\phi)$ appears to be sharper. However, we should stress that some of the models presented in Fig.~\ref{fig:potentials_fractional} predict a value of $V \gg 1$ (in natural units) before the breaking of the perturbative regime (which occurs at $\phi = \phi_{NP}$)! Clearly these models must be rejected as they lead to absurd predictions. \\

\begin{figure}[h]
	\centering
	\includegraphics[width=1.15\linewidth]{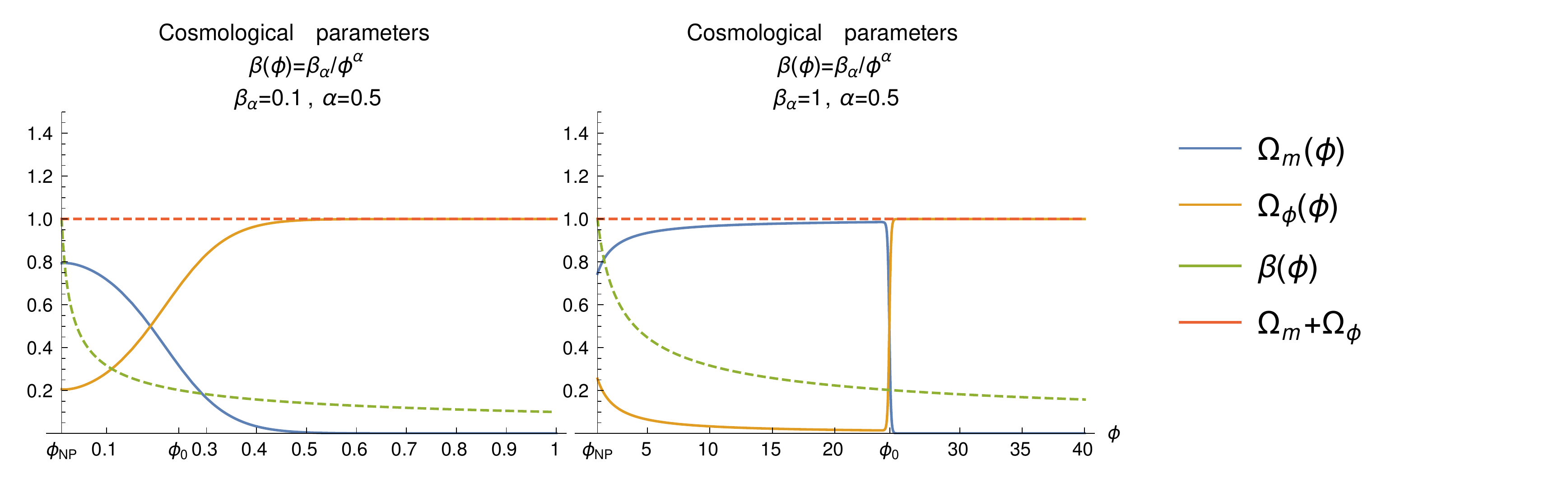}
	\caption{Evolution of the cosmological parameters in the inverse fractional class with $\alpha=0.5$ for $\beta_{\alpha}=0.1$ (left) and $\beta_{\alpha}=1$ (right).}
	\label{fig:comparison_fractional}
\end{figure}

Setting $\alpha=0.5$, in Fig.~\ref{fig:comparison_fractional} we show the evolution of the matter and quintessence normalized energy densities in terms of $\phi$ for two choices of $\beta_{\alpha}$. Similarly to what we have found for the inverse monomial class discussed in Section~\ref{sec:formalism_matter}, in the case $\beta_{\alpha}=0.1$ we do not have $\Omega_m\simeq 1$ (\emph{i.e.} a matter-dominated phase) for $\phi<\phi_0$ before the scalar field drops down to the critical value $\phi_{\mathrm{NP}}$. As a consequence, this model cannot be used to successfully describe quintessence.\\

\begin{figure}[h]
	\centering
	\includegraphics[width=0.6\linewidth]{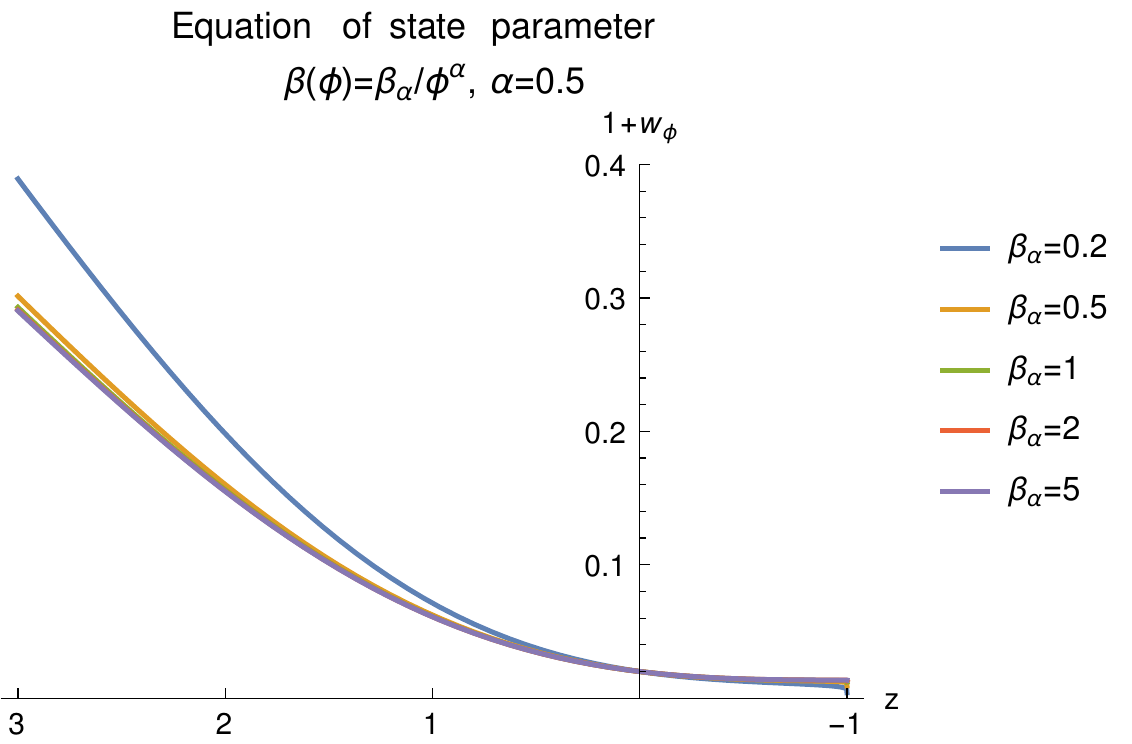}
	\caption{Evolution of the equation of state parameter $w_{\phi}$ versus the redshift $z$ in the inverse fractional class with $\alpha=0.5$ and different values of $\beta_{\alpha}$.}
	\label{fig:all_deviation_050}
\end{figure}

In Fig.~\ref{fig:all_deviation_050} we plot $w_{\phi}$ in terms of the redshift $z$ for $\alpha=0.5$: we notice that, for models with $\beta_{\alpha}\ge 0.5$, the evolution of $w_{\phi}$ does not significantly change.

\subsection{Inverse logarithmic class.}
\begin{figure}[h]
	\centering
	\includegraphics[width=0.6\linewidth]{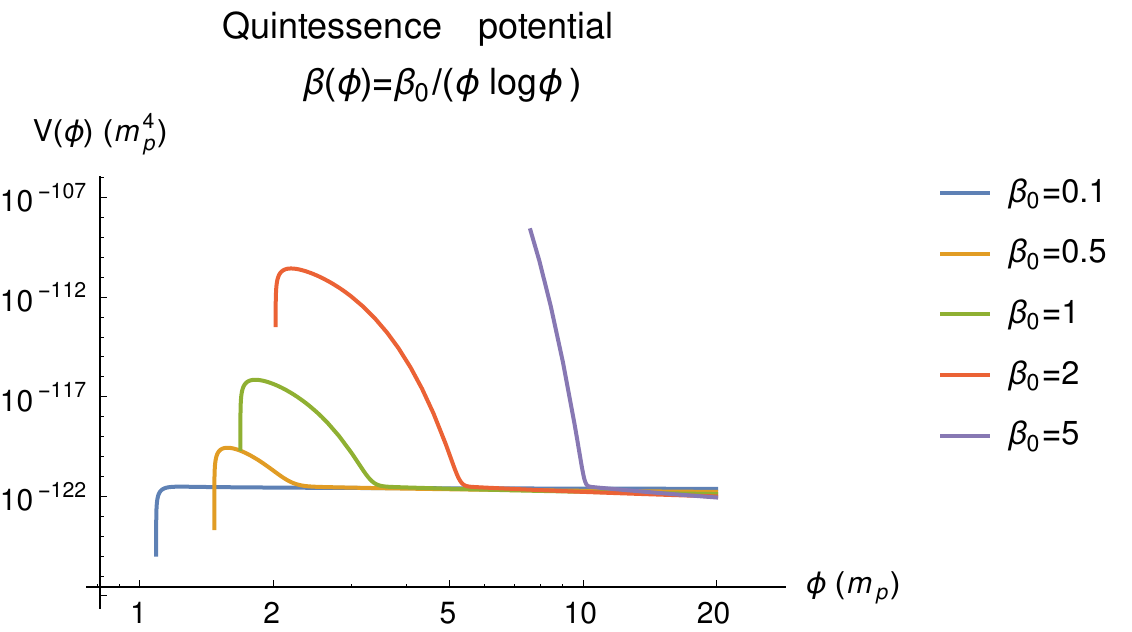}
	\caption{Evolution of the quintessence potential for the inverse logarithmic class for different values of $\beta_0$.}
	\label{fig:all_potentials_ib0}

\end{figure}

In this class the $\beta$-function is given by $\beta(\phi)=\beta_0/(\phi\log\phi) + o(1/\phi\log\phi)$. The quintessence potential is reported in Fig.~\ref{fig:all_potentials_ib0} for different values of $\beta_0$. Even in this case, a transition in $V$ at $\phi\approx\phi_0$ can be clearly appreciated only for values of $\beta_0$ greater than $0.1$. \\

\begin{figure}[h]
	\centering
	\includegraphics[width=1.15\linewidth]{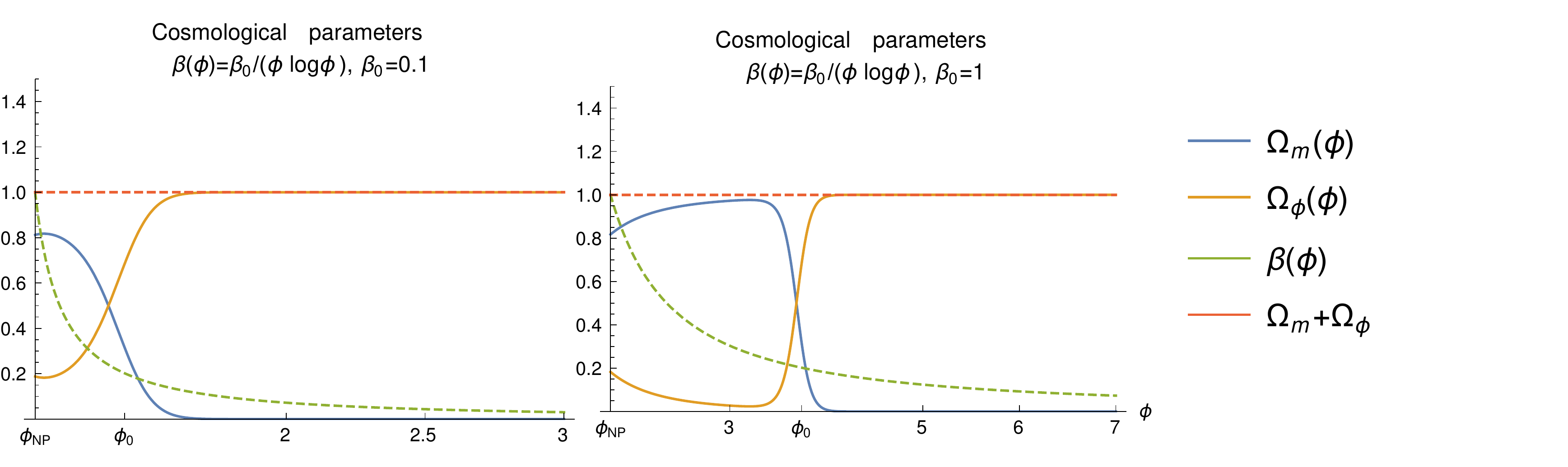}
	\caption{Evolution of the cosmological parameters $\Omega_m$ and $\Omega_{\phi}$ for the inverse logarithmic class with $\beta_0=0.1$ (left) and $\beta_0=1$ (right).}
	\label{fig:comparison_logarithmic}

\end{figure}

In Fig.~\ref{fig:comparison_logarithmic} we show the evolution of the normalized energy densities for the two cases $\beta_0 = 0.1$ and $\beta_0=1$. Consistently with the results shown in Fig.~\ref{fig:all_potentials_ib0}, the matter-dominated phase is correctly reproduced by the model with $\beta_0=1$ but not by the model with $\beta_0=0.1$.

\begin{figure}[h]
	\centering
	\includegraphics[width=0.6\linewidth]{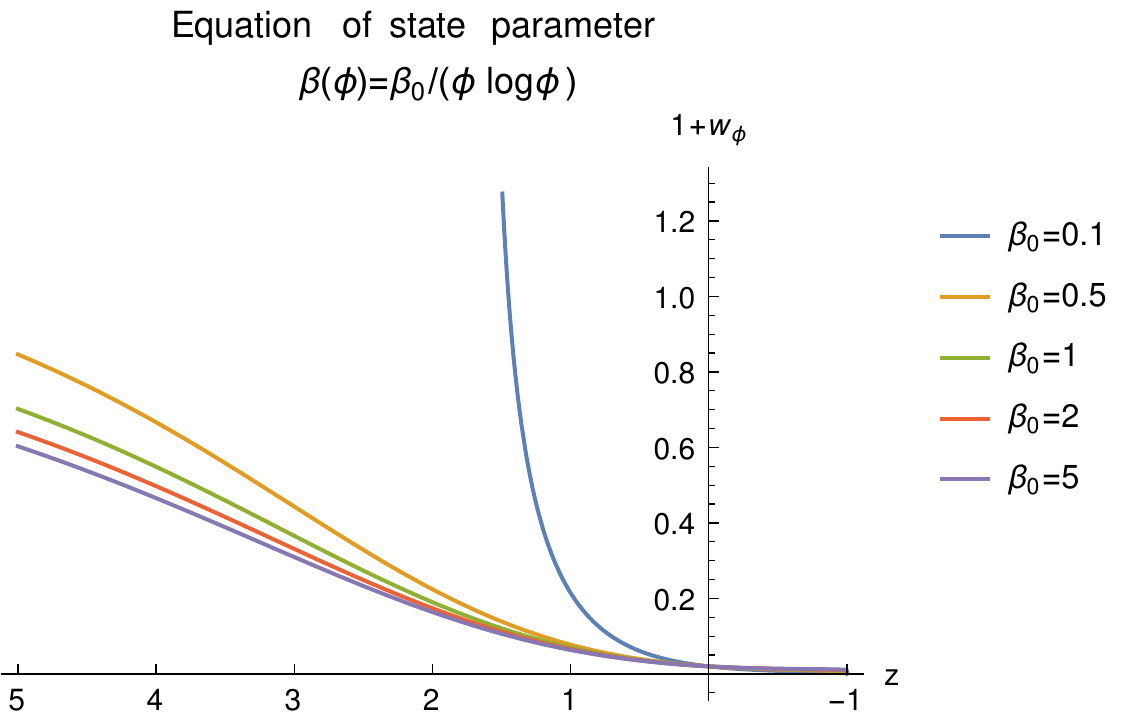}
	\caption{Evolution of the equation-of-state parameter $w_{\phi}$ versus the redshift $z$ in the inverse logarithmic class for different values of $\beta_0$.}
	\label{fig:all_deviation_ib0}

\end{figure}

Finally, in Fig.~\ref{fig:all_deviation_ib0} we show the evolution of the equation-of-state parameter for different values of $\beta_0$. As expected $1+w_{\phi}$ always approaches zero in the infinite future.

\subsection{Power-law class.}
\label{sec:appendix_power_law}
In this section we consider the case $\beta(\phi) = \gamma$ where $\gamma$ is a constant. As explained in Sec.~\ref{sec:univ-quint} (in particular in the footnote~\ref{asymptotic_footnote}), this case is different from the other cases considered in this work as it does not lead to an asymptotic dS solution but rather to an asymptotic power-law solution. Another main difference is that as $\beta$ is exactly constant, after imposing the condition set by Eq.~\eqref{initial} we are not left with any free parameter\footnote{In the following we consider some values of $\gamma$ that, while predicting a value of $w_{\phi0}$ which slightly different from the usual $-0.98$ used in this paper, are still in good agreement with the constraints due to cosmological observations~\cite{Ade:2015xua,Ade:2015rim}.} and the evolution is completely specified! Moreover, as $\beta$ is never becoming of order one, there is no natural definition for a value of $\phi_{NP}$ where the perturbative expansion of $\beta(\phi)$ breaks. In the following we simply stop the evolution at a value of $\phi$ such that we are deep in the matter dominated phase (\emph{i.e.} when $\Omega_m \gg \Omega_{\phi}$). For simplicity and without loss of generality in the following we set $\phi_0 = 0$.\\

As usual we start by presenting, in Fig.~\ref{fig:all_potentials_ib0}, the plots of the evolution of the quintessence potential for different values of $\gamma$. For all these models, the shape of $V$ clearly reflects the transition from matter to quintessence. Moreover, as we have set $\phi_0 = 0$, this transition is always consistently occurring at negative values of $\phi$. \\

\begin{figure}[h!]
	\centering
	\includegraphics[width=0.6\linewidth]{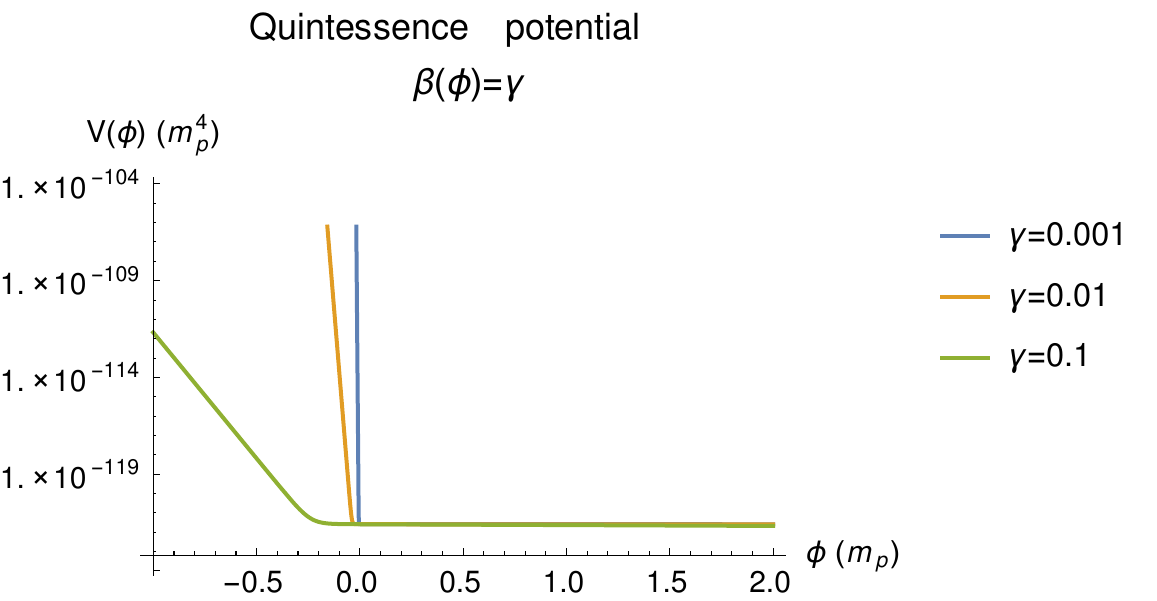}
	\caption{Evolution of the quintessence potential in the power-law class for different choices of $\gamma$.}
	\label{fig:all_potentials_power_rev}
\end{figure}

\begin{figure}[h!]
\begin{minipage}{0.5\linewidth}
	\centering
	\includegraphics[width=\linewidth]{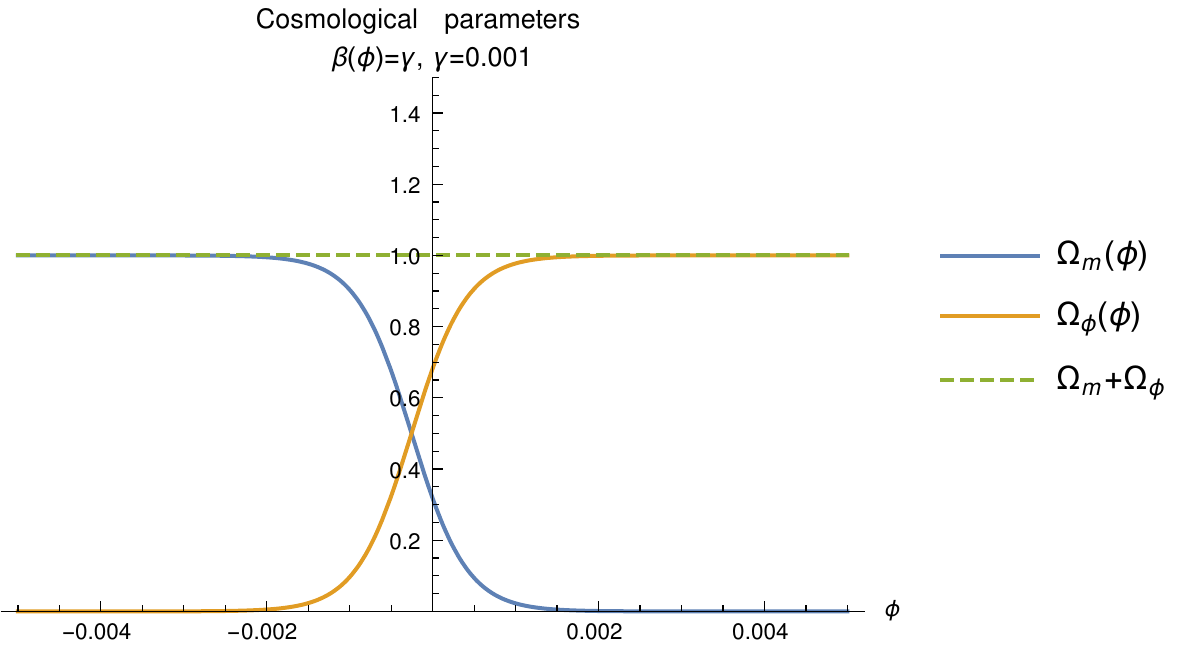}
\end{minipage}%
\begin{minipage}{0.5\linewidth}
	\centering
	\includegraphics[width=\linewidth]{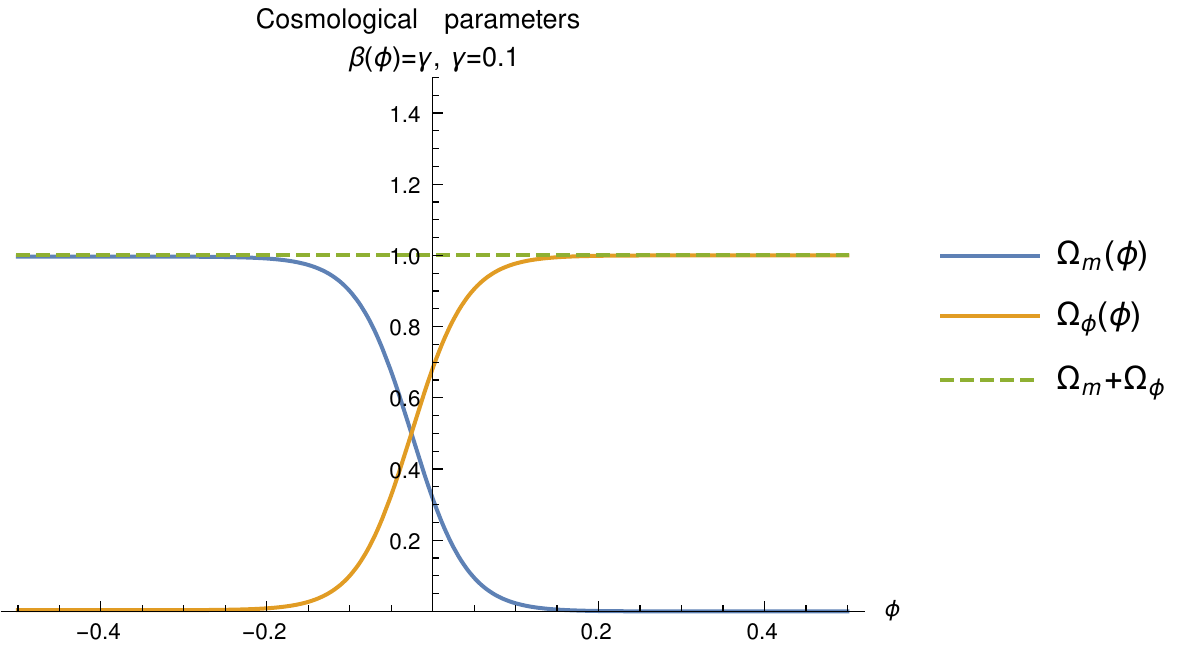}
\end{minipage}
\caption{Evolution of the cosmological parameters in the power-law fractional class for $\gamma=0.001$ (left) and $\gamma=0.1$ (right).}
\label{fig:cosmological_power_rev}
\end{figure}

In Fig.~\ref{fig:cosmological_power_rev}, we present the evolution of the cosmological parameters for $\gamma=0.001$ (left plot) and for $\gamma=0.1$ (right). As it is possible from this figure, the matter dominated phase is correctly reproduced for all these models. \\

\begin{figure}[h!]
	\centering
	\includegraphics[width=0.49\linewidth]{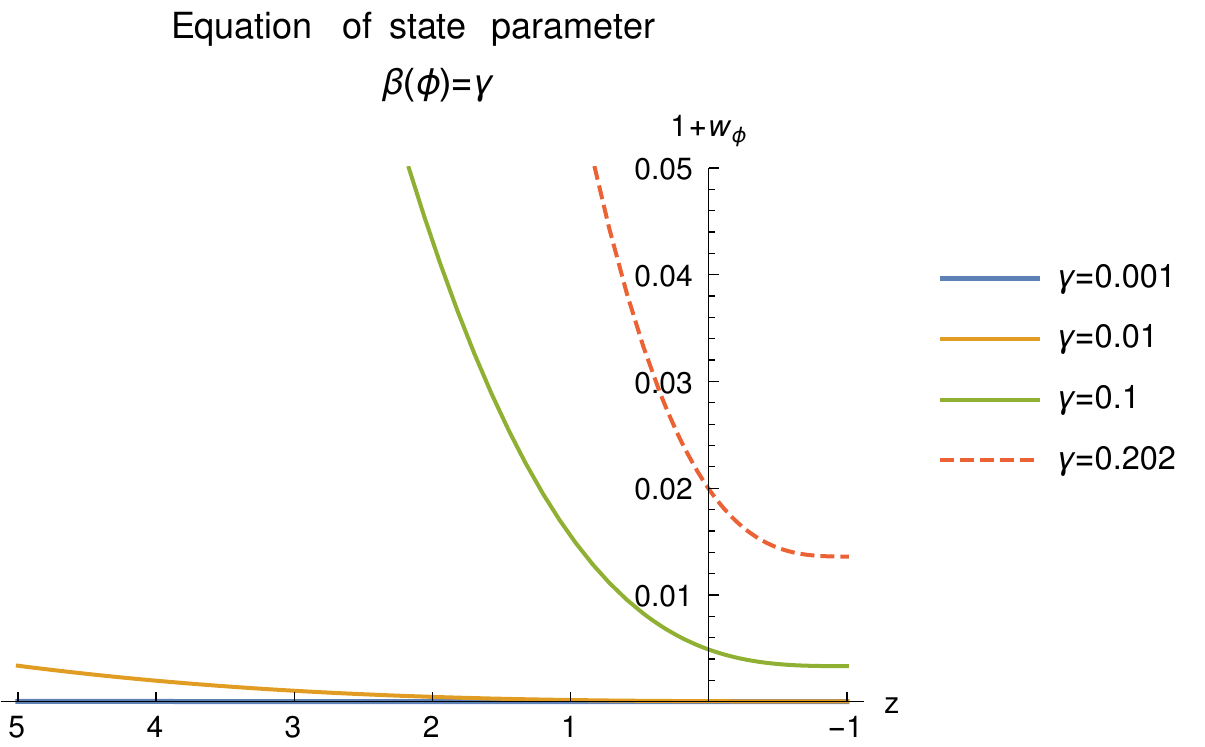} \includegraphics[width=0.49\linewidth]{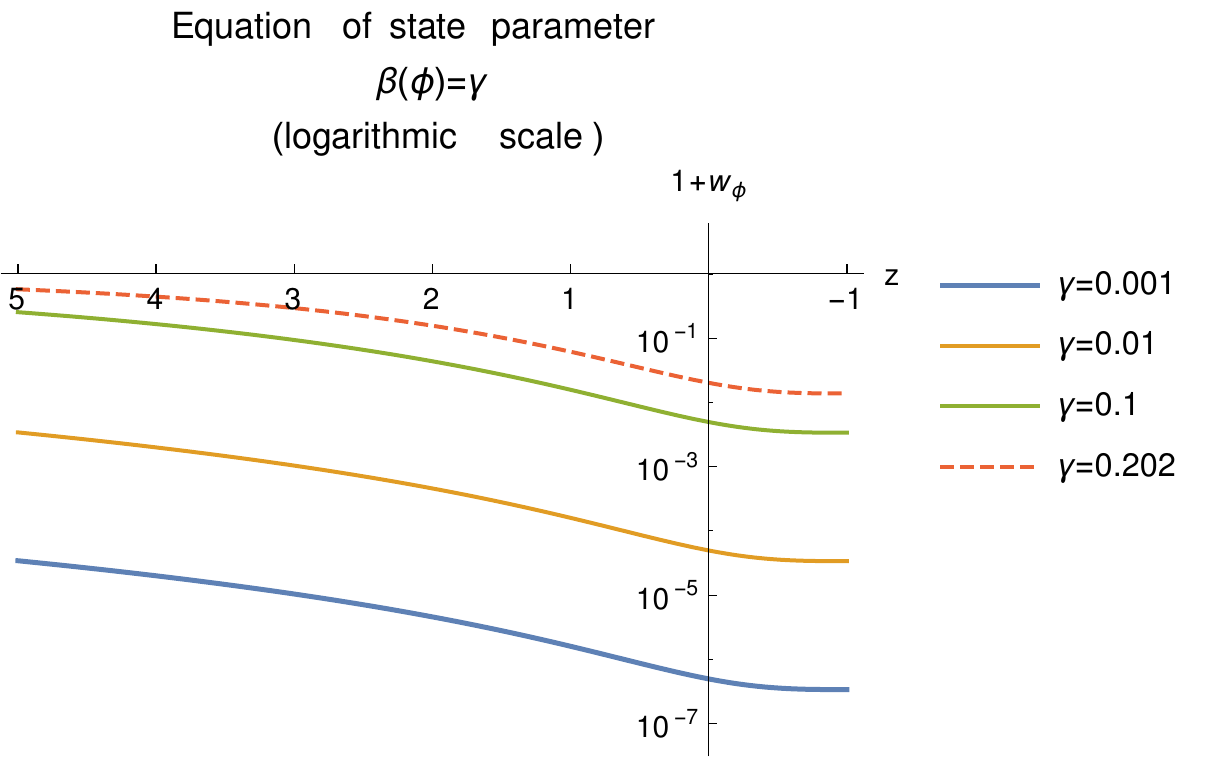}
	\caption{Evolution of the equation-of-state parameter $w_{\phi}$ versus the redshift $z$ in the power-law class for different values of $\gamma$ in linear (left) and logarithmic scale (right).}
	\label{fig:all_deviation_power_rev}
\end{figure}

In Fig.~\ref{fig:all_deviation_power_rev} we show the evolution of the equation of state parameter. In this plot we have also shown (red-dashed line) the evolution for the model which gives $w_{\phi0} = -0.98$. Notice that in the infinite future all these models consistently predict a nearly constant value for $1+w_{\phi}$ which as expected is $1+w_{\phi} \simeq \gamma^2/3$.

\section{Jordan frame metric. \label{esposito}}
In this appendix we show that, as discussed by Esposito-Farese and Polarski in~\cite{EspositoFarese:2000ij}, it is possible to set the metric to be a flat FLRW metric both in the Jordan and in the Einstein frame. If the Einstein frame metric is a flat FLRW, then the metric in the Jordan frame is of the form:
\begin{equation}
\mathrm{d}s^2_J=-\Omega^{-1}(\phi)\mathrm{d}t^2+\Omega^{-1}(\phi)a^2(t)\mathrm{d}\mathbf{x}^2 \ .
\end{equation}
The Einstein equations obtained by varying the action of Eq.~\eqref{jordan} with respect to $\tilde{g}^{\mu\nu}$ then read:
\begin{equation}
\Omega(\phi)\tilde{G}_{\mu\nu}=\tilde{T}_{\mu\nu}^{(\phi)}+\tilde{T}_{\mu\nu}^{(m)} \ ,
\end{equation}
where $\tilde{T}_{\mu\nu}^{(\phi)}$ and $ \tilde{T}_{\mu\nu}^{(m)}$ are respectively given by:
\begin{align}
&\tilde{T}_{\mu\nu}^{(\phi)}=\partial_{\mu}\phi\partial_{\nu}\phi-\tilde{g}_{\mu\nu}\left(\frac{1}{2}\tilde{g}^{\alpha\beta}\partial_{\alpha}\phi\partial_{\beta}\phi+V(\phi)\right) \ , \\
&\tilde{T}_{\mu\nu}^{(m)}=\partial_{\mu}\psi\partial_{\nu}\psi-\tilde{g}_{\mu\nu}\left(\frac{1}{2}\tilde{g}^{\alpha\beta}\partial_{\alpha}\psi\partial_{\beta}\psi+U(\psi)\right) \ ,
\end{align}
so that the only non-zero components of the energy-momentum tensors are:
\begin{align}
&\tilde{T}_{00}^{(\phi)}=\frac{\dot{\phi}^2}{2}+\frac{V}{\Omega(\phi)} \ ,  &\tilde{T}_{ij}^{(\phi)}=a^2(t)\delta_{ij}\left(\frac{\dot{\phi}^2}{2}-\frac{V}{\Omega(\phi)}\right) \ , \\
&\tilde{T}_{00}^{(m)}=\frac{\dot{\psi}^2}{2}+\frac{U}{\Omega(\phi)} \ , &\tilde{T}_{ij}^{(m)}=a^2(t)\delta_{ij}\left(\frac{\dot{\psi}^2}{2}-\frac{U}{\Omega(\phi)}\right) \ .
\end{align}
At this point we can proceed by introducing a new time coordinate $\tau$ and by defining the scale factor $\bar{a}(\tau)$ such that the line element reads:
\begin{equation}
\mathrm{d}s^2_J=-\mathrm{d}\tau^2+\bar{a}^2(\tau)\mathrm{d}\mathbf{x}^2 \ ,
\end{equation}
which implies $\mathrm{d}\tau\equiv \Omega^{-1/2}\mathrm{d}t$ and $\bar{a}(\tau)\equiv \Omega^{-1/2}a(t)$. Under this change of coordinates the stress-energy tensors transform as:
\begin{equation}
\tilde{T}_{\mu\nu}^{(i)}\longrightarrow \tilde{T}_{\mu\nu}^{'(i)}=\frac{\partial x^{\alpha}}{\partial x^{'\mu}}\frac{\partial x^{\beta}}{\partial x^{'\nu}}\tilde{T}_{\alpha\beta}^{(i)} \ .
\end{equation}
Hence:
\begin{align}
& \tilde{T}_{00}^{'(i)}=\pdev{x^{\alpha}}{\tau}\pdev{x^{\beta}}{\tau}\tilde{T}_{\alpha\beta}^{(i)}=\left(\pdev{t}{\tau}\right)^2\tilde{T}_{00}^{(i)}+\pdev{x^i}{\tau}\pdev{x^j}{\tau}\tilde{T}_{ij}^{(i)}=\left(\pdev{t}{\tau}\right)^2\tilde{T}_{00}^{(i)}=\Omega(\phi)\tilde{T}_{00}^{(i)} \ , \\
& \tilde{T}_{ij}^{'(i)}=\frac{\partial x^{\alpha}}{\partial x^i}\frac{\partial x^{\beta}}{\partial x^j}\tilde{T}_{\alpha\beta}^{(i)}=\frac{\partial\tau}{\partial x^i}\frac{\partial\tau}{\partial x^j}\tilde{T}_{00}^{(i)}+\frac{\partial x^k}{\partial x^i}\frac{\partial x^l}{\partial x^j}\tilde{T}_{kl}^{(i)}=\delta^k_i\delta^l_j\tilde{T}_{kl}^{(i)}=\tilde{T}_{ij}^{(i)}\ ,
\end{align}
where we used the fact that, by definition, $\partial\tau/\partial x^i=0, \partial x^i/\partial\tau=0$. Finally, we obtain:
\begin{align}
&\tilde{T}_{00}^{'(\phi)}=\Omega(\phi)\left(\frac{\dot{\phi}^2}{2}+\frac{V}{\Omega(\phi)}\right)=\frac{(\phi')^2}{2}+V\ , \\
&\tilde{T}_{ij}^{'(\phi)}=a^2(t)\delta_{ij}\left(\frac{\dot{\phi}^2}{2}-\frac{V}{f(\phi)}\right)=\bar{a}^2(\tau)\delta_{ij}\left(\frac{(\phi')^2}{2}-V\right)\ , \\
&\tilde{T}_{00}^{'(m)}=\frac{(\psi')^2}{2}+U \ , \\
&\tilde{T}_{ij}^{'(m)}=\bar{a}^2(\tau)\left(\frac{(\psi')^2}{2}-U\right)\ ,
\end{align}
in which we used $\dot{y}(t)=\Omega^{-1/2}y'(\tau)$. These equations show the energy-momentum tensors for $\phi$ and $\psi$ assume the perfect-fluid form in terms of the new variables $\tau,\bar{a}(\tau)$. Therefore, focusing on the matter energy-momentum tensor, both $\tilde{T}_{\mu\nu}^{'(m)}$ in the Jordan frame and $T_{\mu\nu}^{(m)}$ in the Einstein frame can be modeled as
\begin{align}
\tilde{T}_{\mu\nu}^{'(m)}\quad &= \quad(\tilde{\rho}_m+\tilde{p}_m)\tilde{u}_{\mu}\tilde{u}_{\nu}+\tilde{p}_m\tilde{g}_{\mu\nu}, \label{perfect1}\\
T_{\mu\nu}^{(m)}\quad &= \quad (\rho_m+p_m)u_{\mu}u_{\nu}+p_mg_{\mu\nu}, \label{perfect2}
\end{align} 
with $u_i=\tilde{u}_i=0$. From the definition of $T_{\mu\nu}^{(m)}$, we derive the relation between the two tensors:
\begin{align}
T_{\mu\nu}^{(m)} \quad &= \quad -\frac{2}{\sqrt{-g}}\frac{\delta \mathcal{S}_m}{\delta g^{\mu\nu}} \notag \\
&=\quad -\frac{2}{\Omega^2(\phi)\sqrt{-\tilde{g}}}\frac{\delta \mathcal{S}_m}{\delta\tilde{g}^{\alpha\beta}}\frac{\delta \tilde{g}^{\alpha\beta}}{\delta g^{\mu\nu}}, \notag  \\
&=\quad \frac{1}{\Omega^2(\phi)}\tilde{T}^{'(m)}_{\alpha\beta}\cdot \Omega(\phi)\delta^{\alpha}_{\mu}\delta^{\beta}_{\nu}, \notag \\
&= \quad \Omega^{-1}(\phi)\tilde{T}_{\mu\nu}^{'(m)}.
\end{align}
Then, from eqs.~\eqref{perfect1} and~\eqref{perfect2} we obtain
\begin{equation}
(\tilde{\rho}_m+\tilde{p}_m)\tilde{u}_{\mu}\tilde{u}_{\nu}+\tilde{p}_m\tilde{g}_{\mu\nu}=\Omega(\phi)\left[(\rho_m+p_m)u_{\mu}u_{\nu}+pg_{\mu\nu}\right]. \label{perfect3}
\end{equation}
Using the relation between the time coordinates $t$ and $\tau$ it is possible to show that $u_0=\Omega^{1/2}(\phi)\tilde{u}_0$, and hence
\begin{equation}
u_{\mu}u_{\nu}=\Omega(\phi)\tilde{u}_{\mu}\tilde{u}_{\nu},
\end{equation}
which inserted in eq.~\eqref{perfect3} leads to
\begin{equation}
(\tilde{\rho}_m+\tilde{p}_m)\tilde{u}_{\mu}\tilde{u}_{\nu}+\tilde{p}_m\tilde{g}_{\mu\nu}=\Omega^2(\phi)\left[(\rho_m+p_m)\tilde{u}_{\mu}\tilde{u}_{\nu}+p_m\tilde{g}_{\mu\nu}\right].
\end{equation}
We have thus,
\begin{equation}
\rho_m=\Omega^{-2}(\phi)\tilde{\rho}_m,\qquad\qquad p_m=\Omega^{-2}(\phi)\tilde{p}_m,
\end{equation}
or, introducing the function $A(\varphi)\equiv \Omega^{-1/2}(\phi(\varphi))$,
\begin{equation}
\rho_m=A^4(\varphi)\tilde{\rho}_m,\qquad \qquad p_m=A^4(\varphi)\tilde{p}_m,
\end{equation}
which are exactly equal to the ones derived in~\cite{EspositoFarese:2000ij}.

\providecommand{\href}[2]{#2}\begingroup\raggedright


\begin{thebibliography}{10}

\bibitem{Ade:2015xua} 
  P.~A.~R.~Ade {\it et al.} [Planck Collaboration],
  Astron.\ Astrophys.\  {\bf 594}, A13 (2016)
  [arXiv:1502.01589 [astro-ph.CO]].

\bibitem{Starobinsky:1982mr} 
  A.~A.~Starobinsky,
  JETP Lett.\  {\bf 37}, 66 (1983).

\bibitem{Ratra:1987rm} 
  B.~Ratra and P.~J.~E.~Peebles,
  Phys.\ Rev.\ D {\bf 37}, 3406 (1988).

\bibitem{Wetterich:1987fm} 
  C.~Wetterich,
  Nucl.\ Phys.\ B {\bf 302}, 668 (1988).

\bibitem{Peebles:1987ek} 
  P.~J.~E.~Peebles and B.~Ratra,
  Astrophys.\ J.\  {\bf 325}, L17 (1988).

\bibitem{Caldwell:1997ii} 
  R.~R.~Caldwell, R.~Dave and P.~J.~Steinhardt,
  Phys.\ Rev.\ Lett.\  {\bf 80}, 1582 (1998)
  [astro-ph/9708069].

\bibitem{Zlatev:1998tr} 
  I.~Zlatev, L.~M.~Wang and P.~J.~Steinhardt,
  Phys.\ Rev.\ Lett.\  {\bf 82}, 896 (1999)
  [astro-ph/9807002].

\bibitem{Capozziello:2002rd} 
  S.~Capozziello,
  Int.\ J.\ Mod.\ Phys.\ D {\bf 11}, 483 (2002)
  [gr-qc/0201033].

\bibitem{Capozziello:2003tk} 
  S.~Capozziello, S.~Carloni and A.~Troisi,
  Recent Res.\ Dev.\ Astron.\ Astrophys.\  {\bf 1}, 625 (2003)
  [astro-ph/0303041].

\bibitem{Brans:1961sx} 
  C.~Brans and R.~H.~Dicke,
  Phys.\ Rev.\  {\bf 124}, 925 (1961).

\bibitem{Deffayet:2011gz} 
  C.~Deffayet, X.~Gao, D.~A.~Steer and G.~Zahariade,
  Phys.\ Rev.\ D {\bf 84}, 064039 (2011)
  [arXiv:1103.3260 [hep-th]].

\bibitem{Kobayashi:2011nu} 
  T.~Kobayashi, M.~Yamaguchi and J.~Yokoyama,
  Prog.\ Theor.\ Phys.\  {\bf 126}, 511 (2011)
  [arXiv:1105.5723 [hep-th]].

\bibitem{Horndeski:1974wa} 
  G.~W.~Horndeski,
  Int.\ J.\ Theor.\ Phys.\  {\bf 10}, 363 (1974).

\bibitem{Cooper:1982du} 
  F.~Cooper and G.~Venturi,
  Phys.\ Rev.\ D {\bf 24}, 3338 (1981).

\bibitem{Finelli:2007wb} 
  F.~Finelli, A.~Tronconi and G.~Venturi,
  Phys.\ Lett.\ B {\bf 659}, 466 (2008)
  [arXiv:0710.2741 [astro-ph]].

\bibitem{Uzan:1999ch} 
  J.~P.~Uzan,
  Phys.\ Rev.\ D {\bf 59}, 123510 (1999)
  [gr-qc/9903004].


\bibitem{Chiba:1999wt} 
  T.~Chiba,
  Phys.\ Rev.\ D {\bf 60}, 083508 (1999)
  [gr-qc/9903094].


\bibitem{Amendola:1999qq} 
  L.~Amendola,
  Phys.\ Rev.\ D {\bf 60}, 043501 (1999)
  [astro-ph/9904120].

\bibitem{Perrotta:1999am} 
  F.~Perrotta, C.~Baccigalupi and S.~Matarrese,
  Phys.\ Rev.\ D {\bf 61}, 023507 (1999)
  [astro-ph/9906066].

\bibitem{Bertolami:1999dp} 
  O.~Bertolami and P.~J.~Martins,
  Phys.\ Rev.\ D {\bf 61}, 064007 (2000)
  [gr-qc/9910056].

\bibitem{Boisseau:2000pr} 
  B.~Boisseau, G.~Esposito-Farese, D.~Polarski and A.~A.~Starobinsky,
  Phys.\ Rev.\ Lett.\  {\bf 85}, 2236 (2000)
  [gr-qc/0001066].

\bibitem{EspositoFarese:2000ij} 
  G.~Esposito-Farese and D.~Polarski,
  Phys.\ Rev.\ D {\bf 63}, 063504 (2001)
  [gr-qc/0009034].

  

\bibitem{Amendola:2012ys} 
  L.~Amendola {\it et al.} [Euclid Theory Working Group Collaboration],
  Living Rev.\ Rel.\  {\bf 16}, 6 (2013)
  [arXiv:1206.1225 [astro-ph.CO]].


\bibitem{Copeland:1997et} 
  E.~J.~Copeland, A.~R.~Liddle and D.~Wands,
  Phys.\ Rev.\ D {\bf 57}, 4686 (1998)
  [gr-qc/9711068].

\bibitem{Liddle:1998xm} 
  A.~R.~Liddle and R.~J.~Scherrer,
  Phys.\ Rev.\ D {\bf 59}, 023509 (1999)
  [astro-ph/9809272].

\bibitem{Steinhardt:1999nw} 
  P.~J.~Steinhardt, L.~M.~Wang and I.~Zlatev,
  Phys.\ Rev.\ D {\bf 59}, 123504 (1999)
  [astro-ph/9812313].

\bibitem{Binetruy:2014zya} 
  P.~Bin\'etruy, E.~Kiritsis, J.~Mabillard, M.~Pieroni and C.~Rosset,
  JCAP {\bf 1504}, no. 04, 033 (2015)
  [arXiv:1407.0820 [astro-ph.CO]].

\bibitem{Pieroni:2016gdg} 
  M.~Pieroni,
  arXiv:1611.03732 [gr-qc].

\bibitem{Pieroni:2015cma} 
  M.~Pieroni,
  JCAP {\bf 1602}, no. 02, 012 (2016)
  [arXiv:1510.03691 [hep-ph]].

\bibitem{Binetruy:2016hna} 
  P.~Bin\'etruy, J.~Mabillard and M.~Pieroni,
  arXiv:1611.07019 [gr-qc].

\bibitem{Salopek:1990jq} 
  D.~S.~Salopek and J.~R.~Bond,
  Phys.\ Rev.\ D {\bf 42}, 3936 (1990).

\bibitem{Muslimov:1990be} 
  A.~G.~Muslimov,
  Class.\ Quant.\ Grav.\  {\bf 7}, 231 (1990).
  
\bibitem{Maldacena:1997re} 
  J.~M.~Maldacena,
  Int.\ J.\ Theor.\ Phys.\  {\bf 38}, 1113 (1999)
  [Adv.\ Theor.\ Math.\ Phys.\  {\bf 2}, 231 (1998)]
  [hep-th/9711200].

\bibitem{McFadden:2009fg} 
  P.~McFadden and K.~Skenderis,
  Phys.\ Rev.\ D {\bf 81}, 021301 (2010)
  [arXiv:0907.5542 [hep-th]].

\bibitem{McFadden:2010na} 
  P.~McFadden and K.~Skenderis,
  J.\ Phys.\ Conf.\ Ser.\  {\bf 222}, 012007 (2010)
  [arXiv:1001.2007 [hep-th]].
 
\bibitem{Kiritsis:2013gia} 
  E.~Kiritsis,
  JCAP {\bf 1311}, 011 (2013)
  [arXiv:1307.5873 [hep-th]].

\bibitem{Garriga:2014fda} 
  J.~Garriga, K.~Skenderis and Y.~Urakawa,
  JCAP {\bf 1501}, no. 01, 028 (2015)
  [arXiv:1410.3290 [hep-th]].

\bibitem{Garriga:2015tea} 
  J.~Garriga, Y.~Urakawa and F.~Vernizzi,
  JCAP {\bf 1602}, no. 02, 036 (2016)
  [arXiv:1509.07339 [hep-th]].
  
\bibitem{Garriga:2016poh} 
  J.~Garriga and Y.~Urakawa,
  JCAP {\bf 1610}, no. 10, 030 (2016)
  [arXiv:1606.04767 [hep-th]].

\bibitem{Binetruy:2006ad} 
  P.~Bin\'etruy,
  Oxford, UK: Oxford Univ. Pr. (2006) 520 p
  
\bibitem{Capozziello:2005ra} 
  S.~Capozziello, V.~F.~Cardone, E.~Piedipalumbo and C.~Rubano,
  Class.\ Quant.\ Grav.\  {\bf 23}, 1205 (2006)
  [astro-ph/0507438].

\bibitem{Rubano:2001su} 
  C.~Rubano and P.~Scudellaro,
  Gen.\ Rel.\ Grav.\  {\bf 34}, 307 (2002)
  [astro-ph/0103335].

\bibitem{Rubano:2003et} 
  C.~Rubano, P.~Scudellaro, E.~Piedipalumbo, S.~Capozziello and M.~Capone,
  Phys.\ Rev.\ D {\bf 69}, 103510 (2004)
  [astro-ph/0311537].
 
\bibitem{Pavlov:2001dt} 
  M.~Pavlov, C.~Rubano, M.~Sazhin and P.~Scudellaro,
  Astrophys.\ J.\  {\bf 566}, 619 (2002)
  [astro-ph/0106068].

\bibitem{Rubano:2002mc} 
  C.~Rubano and M.~Sereno,
  Mon.\ Not.\ Roy.\ Astron.\ Soc.\  {\bf 335}, 30 (2002)
  [astro-ph/0203205].

\bibitem{Demianski:2004qt} 
  M.~Demianski, E.~Piedipalumbo, C.~Rubano and C.~Tortora,
  Astron.\ Astrophys.\  {\bf 431}, 27 (2005)
  [astro-ph/0410445].
  
\bibitem{Ade:2015rim} 
  P.~A.~R.~Ade {\it et al.} [Planck Collaboration],
  Astron.\ Astrophys.\  {\bf 594}, A14 (2016)
  [arXiv:1502.01590 [astro-ph.CO]].

\bibitem{Khoury:2003rn} 
  J.~Khoury and A.~Weltman,
  Phys.\ Rev.\ D {\bf 69}, 044026 (2004)
  [astro-ph/0309411].
  
\bibitem{Khoury:2013yya} 
  J.~Khoury,
  Class.\ Quant.\ Grav.\  {\bf 30}, 214004 (2013)
  [arXiv:1306.4326 [astro-ph.CO]].

\bibitem{Kallosh:2013xya} 
  R.~Kallosh and A.~Linde,
  JCAP {\bf 1306}, 028 (2013)
  [arXiv:1306.3214 [hep-th]].

\bibitem{Kallosh:2013hoa} 
  R.~Kallosh and A.~Linde,
  JCAP {\bf 1307}, 002 (2013)
  [arXiv:1306.5220 [hep-th]].

\bibitem{Kallosh:2013daa} 
  R.~Kallosh and A.~Linde,
  JCAP {\bf 1312}, 006 (2013)
  [arXiv:1309.2015 [hep-th]].

\bibitem{Kallosh:2013yoa} 
  R.~Kallosh, A.~Linde and D.~Roest,
  JHEP {\bf 1311}, 198 (2013)
  [arXiv:1311.0472 [hep-th]].

\bibitem{Kallosh:2014rga} 
  R.~Kallosh, A.~Linde and D.~Roest,
  JHEP {\bf 1408}, 052 (2014)
  [arXiv:1405.3646 [hep-th]].

\bibitem{Kallosh:2013tua} 
  R.~Kallosh, A.~Linde and D.~Roest,
  Phys.\ Rev.\ Lett.\  {\bf 112}, no. 1, 011303 (2014)
  [arXiv:1310.3950 [hep-th]].

\bibitem{Kallosh:2010ug} 
  R.~Kallosh and A.~Linde,
  JCAP {\bf 1011}, 011 (2010)
  [arXiv:1008.3375 [hep-th]].

\bibitem{Bourdier:2013axa} 
  J.~Bourdier and E.~Kiritsis,
  Class.\ Quant.\ Grav.\  {\bf 31}, 035011 (2014)
  [arXiv:1310.0858 [hep-th]].

\end{thebibliography}
\end{document}